%% file: manuscript.tex
\documentclass[floatfix,aps,prl,superscriptaddress,longbibliography, titlepage, 11pt]{revtex4-1}
\usepackage{subfiles}
\usepackage{amsmath}
\usepackage{enumerate}
\usepackage{graphicx}
\usepackage{color}

\usepackage{mathrsfs}
\usepackage{xcolor}
\mathchardef\mhyphen="2D

\begin{document}

\subfile{main}

\bibliography{reference}

\newpage

\subfile{supplementary}

\end{document}

%% file: main.tex
\title{Artificial electrostatic crystals: a new platform for creating correlated quantum states}

\author{Daisy Q. Wang} 
\affiliation{School of Physics, University of New South Wales, Kensington, NSW 2052, Australia}
\affiliation{Australian Research Council Centre of Excellence in Future Low-Energy Electronics Technologies,The University of New South Wales, Sydney 2052, Australia}

\author{Zeb Krix}
\affiliation{School of Physics, University of New South Wales, Kensington, NSW 2052, Australia}
\affiliation{Australian Research Council Centre of Excellence in Future Low-Energy Electronics Technologies,The University of New South Wales, Sydney 2052, Australia}

\author{Olga A. Tkachenko}
\affiliation{Rzhanov Institute of Semiconductor Physics, Novosibirsk, 630090, Russia}

\author{Vitaly A. Tkachenko}
\affiliation{Rzhanov Institute of Semiconductor Physics, Novosibirsk, 630090, Russia}
\affiliation{Novosibirsk State University, Novosibirsk, 630090, Russia}

\author{Chong Chen}
\affiliation{Cavendish Laboratory, J. J. Thomson Avenue, Cambridge, CB3 0HE, United Kingdom}

\author{Ian Farrer\footnote{Present Address: Department of Electronic and Electrical Engineering, The University of Sheffield, Mappin Street, Sheffield, S1 3JD, United Kingdom.} }  
\affiliation{Cavendish Laboratory, J. J. Thomson Avenue, Cambridge, CB3 0HE, United Kingdom}

\author{David A. Ritchie}
\affiliation{Cavendish Laboratory, J. J. Thomson Avenue, Cambridge, CB3 0HE, United Kingdom}

\author{Oleg P. Sushkov}
\affiliation{School of Physics, University of New South Wales, Kensington, NSW 2052, Australia}
\affiliation{Australian Research Council Centre of Excellence in Future Low-Energy Electronics Technologies,The University of New South Wales, Sydney 2052, Australia}

\author{Alexander R. Hamilton}
\affiliation{School of Physics, University of New South Wales, Kensington, NSW 2052, Australia}
\affiliation{Australian Research Council Centre of Excellence in Future Low-Energy Electronics Technologies,The University of New South Wales, Sydney 2052, Australia}

\author{Oleh Klochan}
\affiliation{School of Physics, University of New South Wales, Kensington, NSW 2052, Australia}
\affiliation{Australian Research Council Centre of Excellence in Future Low-Energy Electronics Technologies,The University of New South Wales, Sydney 2052, Australia} 
\affiliation{School of Science, The University of New South Wales, Canberra, ACT 2612, Australia}

\date{\today}
\maketitle

{\bf The electronic properties of solids are determined by the crystal structure and interactions between electrons, giving rise to a variety of collective phenomena including superconductivity, strange metals and correlated insulators. The mechanisms underpinning many of these collective phenomena remain unknown, driving interest in creating artificial crystals which replicate the system of interest while allowing precise control of key parameters. Cold atoms trapped in optical lattices provide great flexibility and tunability~\cite{Tarruell12, Goldman16}, but cannot replicate the long-range Coulomb interactions and long-range hopping that drive collective phenomena in real crystals. Solid-state approaches support long-range hopping and interactions, but previous attempts with laterally patterned semiconductor systems were not able to create tunable low-disorder artificial crystals, while approaches based on Moir\'{e} superlattices in twisted two-dimensional (2D) materials~\cite{Kennes21,Mak22} have limited tunability and control over lattice geometry. Here we demonstrate the formation of highly tunable artificial crystals by superimposing a periodic electrostatic potential on the 2D electron gas in an ultra-shallow (25 nm deep) GaAs quantum well. The 100 nm period artificial crystal is identified by the formation of a new bandstructure, different from the original cubic crystal and specific to the artificial triangular lattice: transport measurements show the Hall coefficient changing sign as the chemical potential sweeps through the artificial bands. Uniquely, the artificial bandstructure can be continuously tuned to form linear graphene-like and flat kagome-like bands in a single device. A strong insulating state is observed at half filling of the kagome flat band, which is not expected in the absence of strong interactions. This state, unique to the kagome lattice, is consistent with a loop-current Wigner insulator, which arises from long-range Coulomb interaction and delocalised electrons between neighbouring empty sites. 
%We show that this correlated insulating state is consistent with a loop-current Wigner insulator state on the kagome lattice, which can arise due to long-range Coulomb interaction and delocalised electrons between neighbouring empty sites. 
The ability to continuously tune the bandstructure and access flat bands through electrical gating within a single device opens a new route to studying collective quantum states.}

A key challenge in condensed-matter physics is understanding strongly interacting quantum systems where many-body correlated states such as superconductivity emerge. Artificial crystals, in which the key parameters can be controlled in-situ, provide a powerful tool to simulate and study these complex systems. Compared to optical systems, the pioneer in the field of artificial crystals, solid-state based artificial crystals exhibit a pivotal advantage: the easy integration of the long-range Coulomb interactions, which are critical for emulating collective behaviour of real materials. However creating solid state artificial crystals is a non-trivial task. The major challenges lie in fabricating a highly uniform periodic potential $U(r)$ with an amplitude much larger than the Fermi energy $E_F$, while maintaining very low levels of disorder $\Gamma \ll U$.

Most artificial solid state crystals fall into two categories distinguished by the nature of the supperlattice potential. In Moiré superlattices the periodic lattice potential is created by stacking atomically thin 2D materials~\cite{Geim13,Andrei20}. The interaction between different layers can lead to formation of isolated flat bands where a diverse range of correlated electronic phases have been observed~\cite{Cao18,Cao18_2,Lu19,Balents20,Regan20}. %Despite these advances, these Moiré systems face their own set of challenges: the symmetry and strength of the artificial lattice potential are constrained by the host crystal, while twist angle disorder remains a difficult obstacle to overcome~\cite{Uri20}. 
An alternative approach involves artificially patterned superlattices imposed on conventional 2D systems, with the advantage that arbitrary lattice geometries can be created with excellent control. Early studies of superlattices on doped semiconductor heterostructures revealed Weiss oscillations~\cite{weiss1989, Winkler89} and signatures of Hofstadter physics~\cite{Albrecht99, Deutschmann01, Geisler04}, but the weak artificial lattice potential ($U(r) \ll E_F$) and disorder prevented the formation of an artificial solid-state crystal. Recent optical studies of honeycomb lattices etched into GaAs quantum wells~\cite{Gibertini09,Singha11} have revealed characteristics of the honeycomb bands~\cite{Wang_Pinczuk2018} and possible many-body effects~\cite{Du18, Du21}, but etched systems do not allow continuous tuning of the superlattice potential. 
In this work we present a low-disorder two-dimensional artificial crystals defined in semiconductor heterostructures by nanolithographical patterning of electrostatic gates. The flexibility of our design enables continuous tuning to form a graphene-like crystal, or a kagome-like crystal, within a single sample. The observation of an artificial electronic kagome lattice~\cite{Itiro51}, where destructive interference between electron wavefunctions induces an electronic flat band, will allow new studies of a wide range of exotic quantum phenomena~\cite{Balents10} and a rich variety of correlated effects~\cite{Yin22}.

\begin{figure}[h!]
\centering
\includegraphics[width=\textwidth]{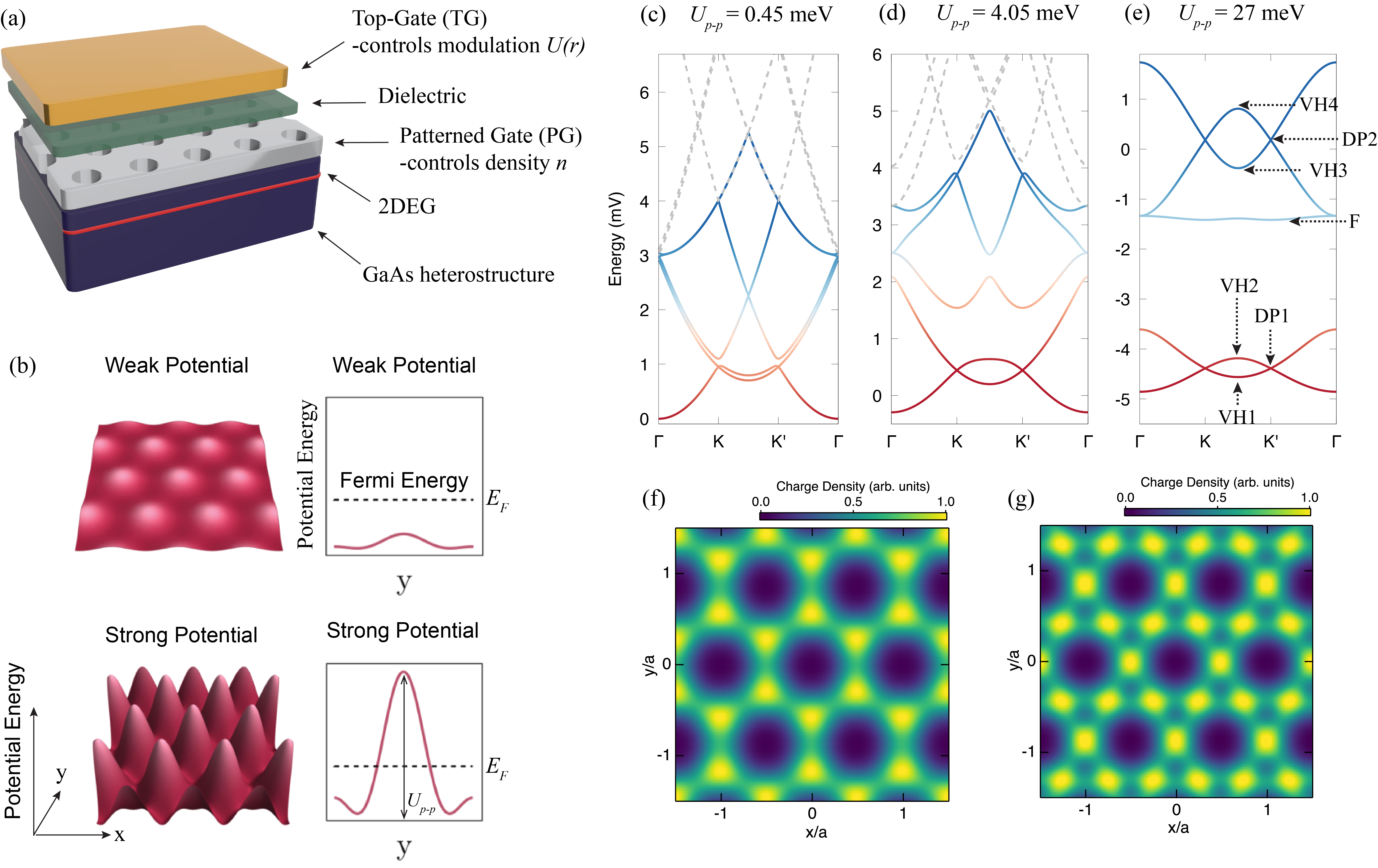}
\caption{{\bf Tunable bandstructure of the electrically defined artificial crystal.} (a) Schematic of the device, showing the double-layer gate design. A surface metal patterned gate (PG, closest to the 2DEG) is patterned with a triangular array of holes and positively biased to induce electrons at the $\textrm{GaAs/Al}$$_{0.6}$$\textrm{Ga}$$_{0.4}$$\textrm{As}$ heterointerface 25 nm below the gate. The lattice constant is $L=100~\textrm{nm}$ and the hole diameter is 45 nm. The overall top gate (TG), separated by a thin $\mathrm{AlO_x}$ dielectric, controls the strength of the superlattice potential through the holes etched in the patterned gate. A strong modulation potential $U(r)$ is essential for creating an artificial crystal with well defined bandstructure. (b) The modulation potential in the weak $U_{p-p}<E_F$ and strong ($U_{p-p}>E_F$) limits. (c, d, e) Evolution of the calculated bandstructure for a 100 nm lattice spacing with different peak-to-peak modulation amplitudes $U_{p-p}$. (c) For a weak modulation potential ($U < E_F$, typically $E_F \sim 2.5$ meV), the mini-bands have mostly parabolic energy dispersion, with small splittings near the artificial Brillouin zone boundary. (d) As the strength of the modulation potential increases the splittings at the zone boundaries become larger, initially only the lowest graphene-like bands are well defined; higher bands overlap and form a `spaghetti' of intersecting levels. (e) Only at very strong potential modulation do the mini-bands become distinct and non-overlapping. In addition to the two graphene-like bands at lower energies (red), there are three kagome-like bands at higher energies (blue). Special points in the bandstructure can be identified in experiments: the van Hove singularities at the band edges (VH1-VH4), the Dirac points at band crossings (DP1 and DP2), and the flat band (F). (f,g) The real-space charge distributions are calculated for the graphene-like bands at strong modulation strength and (g) for kagome-like bands at very strong modulation. For (f), states with energy $-0.3 \mathrm{meV} < \epsilon < \mu=E_F=1.5\mathrm{meV}$ in (d) contribute to the density. For (g), states with energy $-1.5 \mathrm{meV} < \epsilon < \mu=E_F=-0.5 \mathrm{meV}$ in (e) contribute to the density.}
\label{fig:theory}
\end{figure}

In Fig.~\ref{fig:theory}(a) we show a schematic of the dual gate device. Electron beam lithography is used to define a 100 nm period triangular lattice in a metal gate electrode only 25 nm above the GaAs/AlGaAs heterointerface. This electrode defines the lattice geometry and is used to vary the carrier density (band filling). A second overall top gate is deposited on top of a thin dielectric above the patterned gate, and controls the strength of the lattice modulation and thus the artificial bandstructure. The dual gate architecture and the small distance between the patterned gate and the 2D electron gas greatly amplifies the superlattice potential, allowing us to reach the regime where $U(r) \gg E_F$. We eliminate random disorder from dopant atoms by using entirely undoped crystals to ensure $\Gamma \ll E_F, U$ (a comparison of this approach with previous studies is shown in Fig. S1 of Supplementary Section I). 

To calculate the artificial bandstructure we model the lattice as the periodic potential shown in Fig.~\ref{fig:theory}(b)) and described by Eqn.~\ref{uu} in Methods, and perform an exact numerical solution of the single-particle Schr\"odinger equation to obtain the artificial bandstructure (self consistent numerical modelling has also been performed in Refs.~\cite{Tkachenko15,Sushkov13}). The effective strength of the periodic potential is determined by the ratio of the peak-to-peak magnitude $U_{p-p}$ to the Fermi energy $E_{F}$, as sketched in Fig.~\ref{fig:theory}(b). When the superlattice potential is weak the mini-bands are essentially parabolic (Fig.~\ref{fig:theory}(c)), corresponding to nearly-free electrons, a regime that has been studied extensively in experiments with GaAs superlattices~\cite{Albrecht99, Deutschmann01, Geisler04, Wang23}. 
To create an artificial crystal, the superlattice potential must be strong ($U_{p-p} > E_{F}$). In this regime (Fig.~\ref{fig:theory}(d,e)) the energy bands start to separate. Two graphene-like bands at low densities develop first when the modulation strength is strong enough (Fig.~\ref{fig:theory}(d)). Going to even stronger modulation causes three kagome-like bands (blue) to develop at high energies (by `graphene-like' and `kagome-like' we mean that the energy dispersion matches that of the corresponding lattice with nearest neighbour hopping). Charge density distributions calculated with the Fermi energy positioned within the graphene-like bands (Fig.~\ref{fig:theory}(f)) or kagome-like bands (Fig.~\ref{fig:theory}(g)) reveal that electrons form a `graphene-crystal' or `kagome-crystal' around the repulsive triangular antidot lattice potential $U_{p-p}$.

The major difference between the artificial bands in Fig.~\ref{fig:theory}(e) and those in natural materials is the smaller bandwidth $W$ (meV instead of eV) due to the larger lattice constant, $L = 100~\mathrm{nm}$. This necessitates very low levels of disorder but allows the Fermi level to be easily swept through the different bands by tuning the voltage on the patterned gate. 
%The Coulomb interaction energy scale, $e^2/\epsilon L$, is also about 1 meV. This means the strong correlation regime, $U/W>1$, can be readily achieved in the kagome flat band for strong potential modulation $U_{p-p}$.
The artificial bandstructure and its topology can be detected through the dynamics of electrons in this artificial crystal. If the Fermi surface expands with increasing energy, the particles in the band are electron-like, whereas if the Fermi surface shrinks then charge carriers are hole-like. Transitions between electron-like and hole-like dynamics can only occur at well defined points in the band structure, namely van Hove (VH) singularities or Dirac points (DP), as labelled in Fig.~\ref{fig:theory}(e). A change of carrier type will result in a change in the sign of the Hall coefficient ($R_H$) with $R_H<0$ indicating electron-like carriers and $R_H>0$ indicating hole-like carriers. This provides a clear experimental signature in electrical transport measurements.

%This is figure 2
\begin{figure}[h!]
\centering
\includegraphics[width=0.9\textwidth]{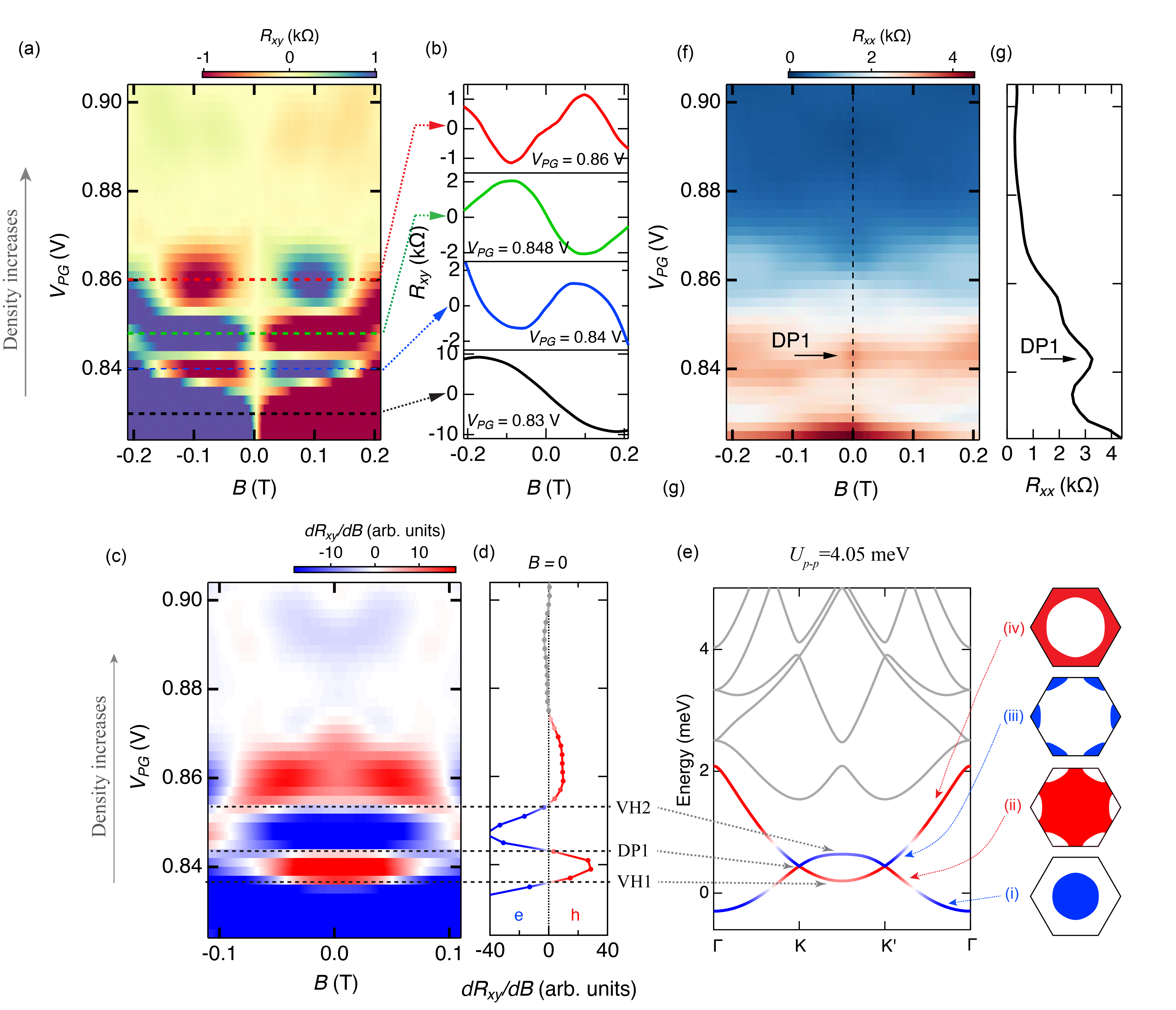}
\caption{{\bf Measuring artificial bandstructure in graphene-like bands}.
  (a) Measured low-field Hall resistance $R_{xy}$ of device D251 (Van der Pauw geometry with $\sim 2900$ lattice sites) at $V_{PG}=-0.5$ V, $T=1.5$ K. (b) Line cuts of the Hall resistance at four different carrier densities indicated by the dashed lines in (a). The Hall slope changes sign near $B = 0$: the black and green lines have a negative slope indicative of electron-like carriers, while the red and blue lines have a positive slope indicative of hole-like carriers. The transition points between electron-like and hole-like behaviour are highlighted in panel (c), which plots the Hall coefficient $R_{H} = dR_{xy}/dB$. Electron-like and hole-like behaviour is illustrated by the colour (blue and red, respectively). Three transition points separate four regions of different carrier type, alternating between electron-like, hole-like, electron-like and hole-like as $V_{PG}$ is increased. A line cut of the Hall coefficient at $B=0$ is plotted in (d) with electron-like (negative) and hole-like (positive) region colored with blue and red respectively. This sequence of transitions is consistent with the calculated band structure (panel (e)), where the two graphene-like bands at low energy have three transition points: VH1, DP1, and VH2. The electron-like (blue) and hole-like (red) behaviour on either side of a transition is colour-coded, with sub panels (i-iv) showing the corresponding Fermi surfaces. (f) Additional evidence for the formation of a Dirac point comes from the longitudinal resistance $R_{xx}$, plotted as a function of $V_{PG}$ and magnetic field, $B$.  A line cut of $R_{xx}$ at $B=0$ in panel (g) shows a clear resistance peak at the position of the Dirac point DP1 (indicated by the black arrow). 
}
\label{fig:dirac}
\end{figure}

Experimentally we probe the formation of the artificial bandstructure using measurements of the low-field magnetoresistance at $T=1.5~\textrm{K}$ while continuously varying $V_{PG}$ (Fig.~\Ref{fig:dirac}). This positive bias on the patterned gate is linearly proportional to the carrier density (band filling) as our device essentially functions as a field-effect transistor (Supplementary section X). As shown in Fig.~\ref{fig:dirac}(a) and highlighted by the linecuts in Fig.~\ref{fig:dirac}(b), the slope of the Hall resistance near $B=0$ undergoes a series of sign changes: from negative to positive, back to negative, and then to positive again as the carrier density increases. This behaviour is further illustrated by calculating the Hall coefficient $R_H=dR_{xy}/dB$ (Fig.~\ref{fig:dirac}(c)), where the linecut at $B=0$ (Fig.~\ref{fig:dirac}(d)) shows switching between electron-like carriers ($R_H<0$) and hole-like carriers ($R_H>0$). In the artificial bandstructure, when the Hall coefficient changes from negative to positive, the carrier type transitions from electron-like to hole-like, indicating the presence of a van Hove singularity. On the other hand, when the Hall coefficient changes from positive to negative, the carrier type transitions from hole-like to electron-like, signifying the presence of a Dirac point. This sequence of varying carrier types from electron-like to hole-like, back to electron-like and then to hole-like again aligns precisely with the calculated bandstructure at a modulation strength of $U_{p-p}=4.05$ meV (Fig.~\ref{fig:dirac}(e)), where the three transition points at which $R_H$ crosses zero are consistent with the positions of VH1, DP1, and VH2 in the calculated band diagram. For high densities (above the last hole-like region, $V_{PG}>0.87$ V), $R_H$ is strongly suppressed as the  mini-bands merge together into a continuum at high energies $E_F \sim U_{p-p}$. We use this point at which $R_{H}$ is suppressed to estimate the value of $U_{p-p}$ (Supplementary section VIII).

%When the superlattice potential is weak the system behaves as a conventional 2DEG. In this well-explored regime~\cite{weiss1989, Wang23} the Hall resistance is featureless over the entire range of 2DEG density, with a negative Hall slope $R_{H} = dR_{xy}/dB$ indicating that charge carriers are electron-like (Fig. S2 of Supplementary section II). When superlattice modulation is large enough, the formation of the artificial crystal modifies the Hall effect: the carriers change behaviour from electron-like to hole-like as the Fermi energy is swept through the artificial bands, causing a change in sign of the Hall coefficient as $V_{PG}$ is increased. The observation of a positive Hall slope is an unambiguous signature of hole-like carriers, which can only arise due to the formation of mini-bands in the artificial crystal. 

%The sign of the Hall slope probes whether charge carriers are electron-like or hole-like, and can hence detect the formation of artificial bands.  We highlight that the results are highly reproducible, not only between devices made on the same heterostructure, but even between devices made with different lattice constants and from different wafers (Supplementary section IV).

Additional evidence for the formation of graphene-like bands comes from the evolution of the longitudinal resistance $R_{xx}$ as a function of carrier density, controlled by $V_{PG}$ (Fig.~\ref{fig:dirac}(f) and (g)). There is a clear resistance peak at $V_{PG}=0.843$ V centered at the position of the Dirac Point DP1, which is expected for a graphene-like system. We estimate the mobility of the charge carriers near the artificial Dirac cone using the resistance and effective density at VH1 or VH2 ($0.6\times 10^{10}~\textrm{cm}^{-2}$) to be $\mu_{D} \approx 100,000 ~\mathrm{cm^2/Vs}$, which is approximately $10$ times higher than that of the host 2DEG at the same carrier density. This high mobility is consistent with the `massless' nature of Dirac carriers in the graphene-like bands. We use the width of the Dirac peak to estimate the disorder, $\Gamma \lesssim 0.1~\textrm{meV}$, which is 40 times smaller than the superlattice potential $U_{p-p}$. 

%Previous studies of gate defined artificial lattices were unable to reach the strong modulation regime $U_{p-p} \sim E_F$ without the system breaking up into 'puddles' due to long range disorder~\cite{liang1994,melinte2004}. Optical studies of honeycomb lattices etched into GaAs quantum wells have shown signatures of linearly dispersing bands~\cite{Wang_Pinczuk2018}, but etched systems do not allow continuous tuning of the superlattice potential. In contrast, here we have tuned continuously from an unperturbed system with massive electrons in parabolic bands to electrons and holes in an artificial graphene bandstructure. 

\begin{figure}[h!]
\centering
\includegraphics[width=0.95\textwidth]{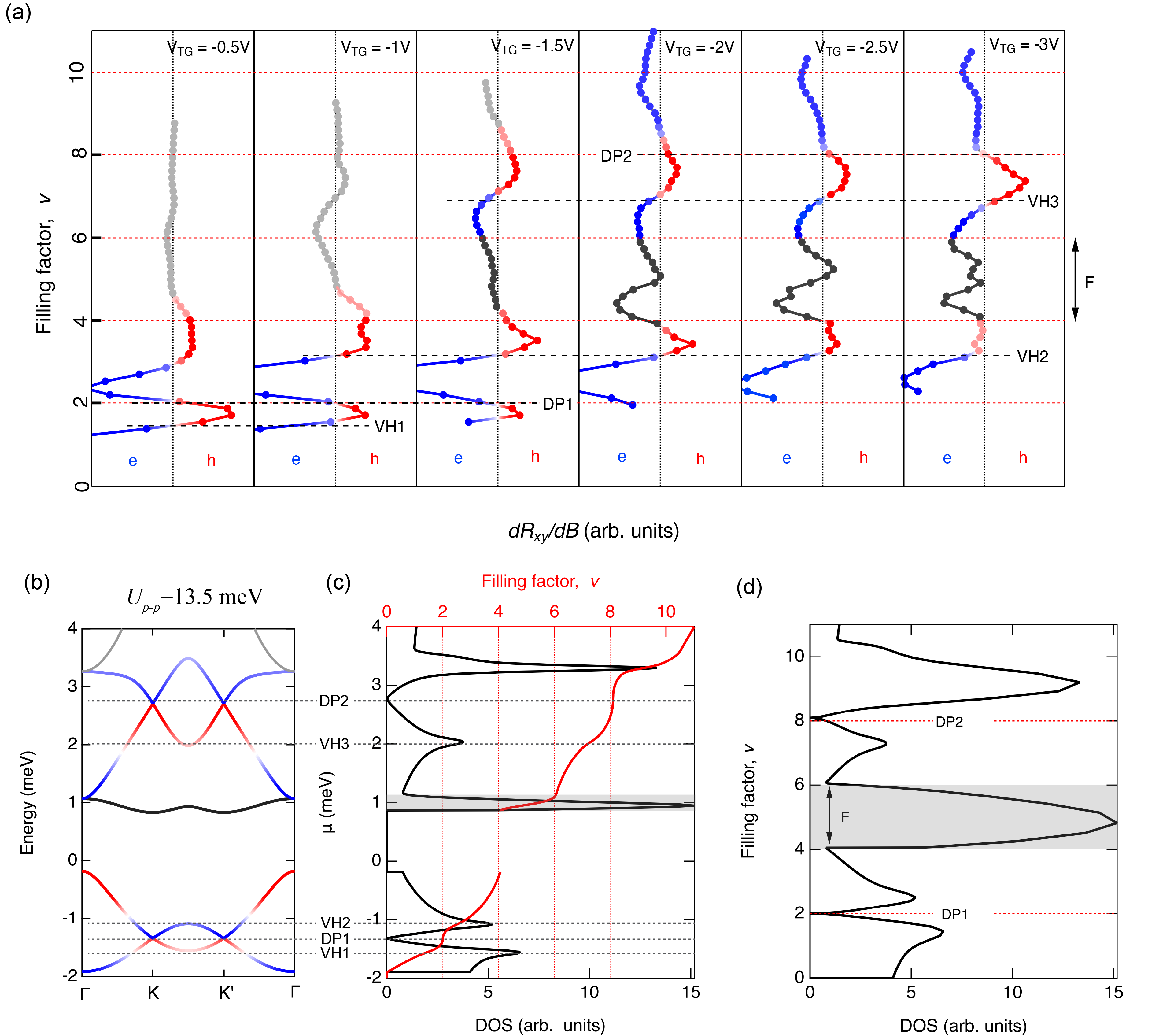}
\caption{{\bf Tuning of the bandstructure and kagome-like bands}. (a) Measured low-field Hall coefficient of device D251 at different modulation strengths, plotted as a function of the filling factor (number of electrons per unit cell). From left to right, the superlattice modulation potential is increased as $V_{PG}$ becomes more negative. Electron-like (negative) and hole-like (positive) Hall coefficient $R_H=dR_{xy}/dB$ are colored as blue and red respectively. Black dashed lines indicate the positions of the Van Hove singularities VH1, VH2 and VH3 where the Hall coefficient crosses zero from negative to positive, as well as Dirac points DP1 and DP2 (corresponding to 1 and 4 full-filled bands) where the Hall coefficient crosses zero from positive to negative. The section corresponding to populating the kagome flat band F ($4<\nu<6 $) is colored black. In this section the Hall coefficient is theoretically not well defined due to the narrow bandwidth. (b) The calculated bandstructure for $U_{p-p}=13.5~\mathrm{meV}$ shows both graphene-like and kagome-like bands (see Supplementary Section VII and VIII for details of the estimation of the superlattice potential strength and the effects of Coulomb screening due to filling of multiple bands). Electron-like and hole-like sections of the bandstructure are color coded with transitions points labeled. Black lines in panels (c) and (d)  show theoretical density of states (DOS= $dN/d\epsilon$) versus chemical potential (c) and versus filling factor (d). The red line in panel (c) shows the chemical potential versus the filling factor. Grey shaded areas in (c) and (d) correspond to filling the kagome flat band F. %(c) Chemical potential corresponding to (b) as a function of DOS on the bottom axis and filling factor $\nu$ on the top axis. Because DOS is zero in the bandgap region from 0 to 1 meV, electron filling jumps directly into the band above. (d) DOS plotted as a function of filling factor and corresponding chanrge in $V_{PG}$ on the right axis. In real experiment, $V_{PG}$ is varied to directly tune the density and thus band filling in the system.
}
\label{fig:kagome}
\end{figure}

Having shown that we can use the electrostatic gate to transform 2D electrons in the GaAs quantum well into a graphene-like crystal, we now exploit the ability to tune the strength of the superlattice potential to create a kagome lattice. Fig.~\ref{fig:kagome}(a) shows the evolution of the Hall coefficient at $B=0$ as the modulation strength is increased by applying a more negative voltage to $V_{TG}$ (complete maps of the Hall coefficient are shown in Fig. S3.) Here we plot $R_H$ as a function of the filling factor $\nu$, the number of electrons per lattice unit cell (which is directly proportional to $V_{PG}$). Because of spin, a full band has a capacity of two electrons per unit cell. The filling factor is calibrated from the spacing of the Hall sign changes, as detailed in Supplementary Section III. This filling factor assignment was verified through measurements of multiple devices on different wafers. To track the evolution of the bandstructure with $V_{TG}$, we follow the experimentally determined transition points that separate the electron-like and hole-like regions, as marked in Fig.~\ref{fig:kagome}(a). 
We found that as $V_{TG}$ is made more negative, the increased modulation causes new sign changes of $R_H$ to emerge at higher band fillings, while the previously identified sign changes associated with the VH1, VH2 and DP1 points in the lower graphene-like bands become suppressed. The order and spacing of these new $R_H$ transitions are fully consistent with the expected formation of kagome-like bands as calculated in Fig.~\ref{fig:kagome}(b), revealing the third van Hove singularities VH3, as well as the second Dirac point DP2.

%The conversion between $\Delta_{PG}$ to electron filling factor $\nu$ is extracted from the end of the hole-like region in the second band to DP2, which corresponds to the exact filling of two bands with 2 electrons per band from spin. We emphasize that this conversion value $\Delta_{PG} \approx 24~\mathrm{mV}$ is only related to the wafer depth (when lattice constant is kept constant) as the device can be modelled by a simple parallel plate capacitor. Multiple devices on various wafer depths and lattice constants have been measured to justify this observation (Supplementary section III). A second verification of the band filling is also provided by the Hofstadter patterns in high field in supplementary section IV.
%From the top of the graphene-like bands to DP2 corresponds to filling two bands, which is used to extract the width of a single band $\Delta V_{PG} \approx 0.024~\mathrm{V}$. This provides a calibration that will later be used to count how many artificial bands are filled for a given change in $V_{PG}$, and to locate the flat band region.
 
In addition to the new van Hove and Dirac points, the formation of kagome-like bands at the stronger modulation offers two more interesting features: one is the opening of a band gap between the kagome-like bands at higher energy and the Dirac-like bands at lower energy; the other is the formation of a kagome flat band with a narrow bandwidth ($\sim 0.2$ meV), where strong correlation effects are expected. The Coulomb interaction energy scale, $e^2/\epsilon L$, is about 1 meV. This means the strong correlation regime can be readily achieved in the kagome flat band when the potential modulation is strong. 
To fully understand the magneto-transport properties of the device in this strong modulation regime, we plot the density of states (DOS=$dN/d\epsilon$) corresponding to Fig.~\ref{fig:kagome}(b) versus chemical potential in Fig.~\ref{fig:kagome}(c) and versus the filling factor in Fig.~\ref{fig:kagome}(d). In the absence of disorder, there are no states within the band gap, so the chemical potential jumps over the band gap as shown by the red line in Fig.~\ref{fig:kagome}(c). This is why the band gap collapses to a single point at $\nu=4$ in Fig.~\ref{fig:kagome}(d). In reality, due to the existence of disorder and impurities, there are some states within the band gap. In our device, based on the density calibration, we estimate the disorder-related capacity of the band gap is less than $5\%$ of the capacity of a band (Supplementary Section X). 
Above the band gap, the DOS shows a strong spike in the kagome flat band(Fig.~\ref{fig:kagome}(c)), which could host possible correlated states. 
(There is another spike in the DOS at $\mu \sim 3.3$ meV, however, because of the existence of both dispersing and non-dispersing electrons in this region, no correlated states are expected.)
Despite the narrow bandwidth in energy ($\sim 0.3$ meV), the flat band holds the same electron density as any other band, and has a filling factor of $\nu = 2$. This is illustrated in Fig.~\ref{fig:kagome}(d) which directly matches the transport measurement when the band filling is varied by $V_{PG}$. %In this single particle picture, DOS is low at the band edge of the flat band and high in the centre of the band. Experimentally this indicates a lower resistance in the centre of the flat band without any correlation effects.

\begin{figure}[h!]
\centering
\includegraphics[width=\textwidth]{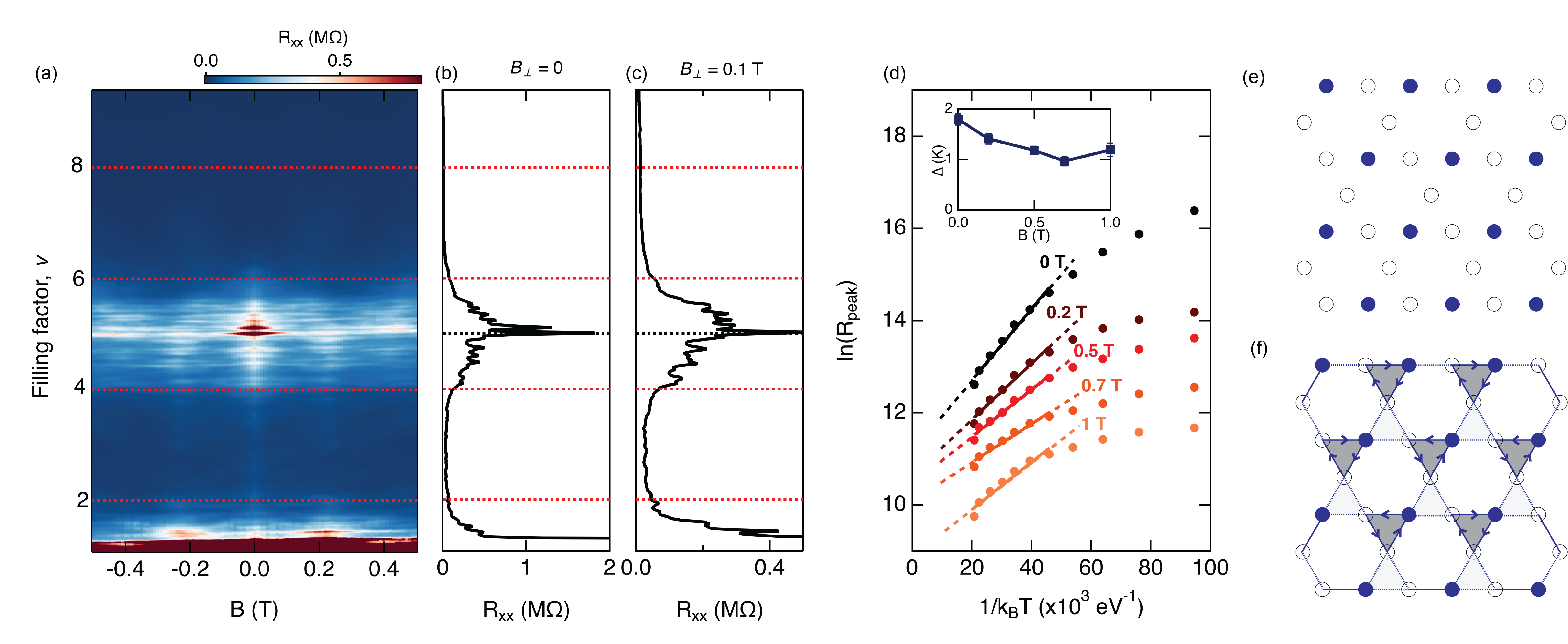}
\caption{{\bf Insulating state in the flat band and loop-current Wigner insulator}. (a) Longitudinal resistance $R_{xx}$ of device D252 (Hallbar geometry with $\sim 600$ lattice sites) with $V_{PG}=-2$ V at $T=350$ mK. Red dashed lines indicate the band filling. Line cuts of $R_{xx}$ at different magnetic field $B_{\perp}=0$ (b) and $B_{\perp}=0.1$ T (c) with black dashed lines indicate half filling of the flat band. (c) Arrhenius plot (circles) of the resistance of the half-filling insulating state at different $B_{\perp}$. The solid lines (with dashed extrapolation) show the fitting with $\exp{[ \,-\Delta/(2k_BT)] \,}$. The inset shows the extracted thermal activation gap $\Delta$ as a function of $B_{\perp}$. (e, f) Schematic of the generalized Wigner insulator and loop-current Wigner insulator state at half filling of the flat band (1/3 filling of the kagome lattice). Solid and dashed lines in (f) illustrates two possible scenarios of the triangular loop current. Arrows on the solid triangles indicating the direction of the loop currents, which is disordered at the measurement temperature in the absence of $B_{\perp}$.
}
\label{fig:flatband}
\end{figure}

In Fig.~\ref{fig:flatband}(a) we show the longitudinal resistance $R_{xx}$ measured on device D252, where a Hall bar geometry is used to obtain more accurate measurements of $R_{xx}$. At very low densities (low band filling factor $\nu$) the sample is insulating due to disorder and Anderson localisation. Increasing the carrier density causes a sharp reduction in $R_{xx}$, as increasing the carrier density screens the static disorder. There is a significant rise in $R_{xx}$ when the kagome flat band starts to fill ($4<\nu<6$), with a very sharp resistance spike at half filling of the flat band. 
%The existence of this peak can not be explained simple disorder and unambiguously points towards a correlated state. 
%The resistance remains high throughout the flat band, and drops sharply once the flat band is filled ($\nu>6$). 
This huge resistance spike cannot be explained by trivial disorder, since disorder would be most significant at the band edges, at $\nu=4$ and $\nu=6$. In contrast the resistance peak occurs at half filling of the flat band, which unambiguously points towards a correlated insulating state. The resistance of this insulating state is extremely sensitive to an out-of-plane magnetic field: it is strongly suppressed by a field of only $B_{\perp}=100\textrm{mT}$, as shown in (Fig~\ref{fig:flatband}(c)). Interestingly, even though the resistance peak is suppressed by a small $B_{\perp}$, thermal activation measurements show that the the size of the energy gap is almost unaffected by the field, staying around $1$ K up to $B_{\perp}=1\textrm{T}$.

%The resistance $R_{xx}$ at $B=0$ shows a strong peak at half filling of the flat band. %Based on our density calibration (Supplementary section IV), the deviation of the peak position from the band centre, the deviation does not exceed $\pm 8\%$ of the band width as indicated by the rose shaded area in Figs~\ref{fig:flatband}.
%The existence of this peak can not be explain by simple disorder and unambiguously points towards a correlated insulating state. We found that the the resistance of this insulating state is extremely sensitive to the out-of-plane magnetic field $B_{\perp}$. Temperature dependence of the resistance peak (Fig~\ref{fig:flatband}(d)) shows a thermal activation gap of $1 - 2$ K. Even though the  This insulating state at half filling of the kagome flat band corresponds to 1/3 sites of the kagome lattice filled, which can not be described by a Mott insulator with only onsite repulsion. Insulating states at fractional fillings of the lattice have previously been associated with a commensurate Wigner insulator state~\cite{Regan20, Krix22} as shown in Fig.~\ref{fig:flatband}(e). However, this commensurate Wigner insulator cannot explain the observed strong dependence on $B_{\perp}$. 

This insulating state at half filling of the kagome flat band corresponds to 1/3 of the sites of the kagome lattice being occupied, which cannot be described by a Mott insulator with only onsite repulsion. Insulating states at fractional fillings of triangular lattices in Moir{\'e} systems have previously been associated with commensurate Wigner insulator states~\cite{Regan20,Xu2020}. However, such commensurate Wigner insulators on a kagome lattice, as illustrated in Fig.~\ref{fig:flatband}(e), cannot explain the observed strong dependence of $R_{xx}$ on $B_{\perp}$. A unique feature of the kagome lattice is that, in the commensurate Wigner insulator configuration, each electron is surrounded by empty sites with only one site occupied per kagome triangle (Fig.~\ref{fig:flatband}(e)). This special configuration at 1/3 filling facilitates electron delocalisation across three neighbouring sites within a kagome triangle. Theoretically, this process is energetically favourable because it does not affect the Coulomb energy, but reduces the zero-point kinetic energy. %There are two possible scenarios of this delocalization which are shown as the solid and dashed lines in Fig.~\ref{fig:flatband}(f) and hence the insulating state is double degenerate. 
This electron delocalisation inevitably leads to a circulating current around the kagome triangle, conceptually resembling the loop current model proposed for cuprates~\cite{Varma97}. 

%This 'loop current Wigner insulator' model can also explain the apparent contradiction between the magnetic field dependence of the resistance and the activation gap. 
The insulating state can be visualized as a series of in-plane loop currents on a triangular lattice as illustrated in Fig.~\ref{fig:flatband}(f). Since each loop current carries a magnetic moment, the insulating state can also be interpreted as a set of orbital magnetic moments on a triangular lattice perpendicular to the plane (Ising type). The magnetic moment per loop is rather large, estimated to be $ \mu_{\pm} = \pm \frac{etL^2}{16} \approx \pm 10\mu_B $ where $L=100$ nm is the lattice constant and $t \approx 0.6$ meV is the nearest neighbour hopping for the tight binding kagome model. The value of $t$ is estimated based on the bandstructure shown in Fig.~\ref{fig:kagome}(b), with details provided in supplementary section XI. In a perfect tight binding kagome model with long-range repulsion but no long-range hopping, the Ising orbital magnetic moments remain disordered down to $T=0$. However, in reality, a small next neighbour hopping term $t'\sim 0.07$ meV induces a very weak antiferromagnetic interaction between orbital magnetic moments $J_{AF} \propto t'^2 \sim 10 - 20$ mK (Supplementary Section XI). As a result the system remains orbitally paramagnetic at experimentally accessible temperatures (we estimate our electron temperature is $\sim 100\textrm{mK}$).

This `loop current Wigner insulator' model also explains the strong suppression of the resistance despite the activation gap being largely unaffected by the magnetic field. The large magnetic moment of the loop currents makes them easily orderable, even under the influence of a tiny magnetic field $B_{\perp}$. Conductivity of the correlated state is only provided by electrons thermally excited over the correlation induced energy gap $\Delta$. At $B_{\perp}=0$, these electrons scatter due to exchange interaction from the thermal fluctuations of the disordered magnetic moments. Applying $B_{\perp}$ orders the loop currents, reducing scattering and thereby increasing conductivity, but does not change the gap. This mechanism agrees well with the experimental observation that resistance decreases with the application of a small $B_{\perp}$, while the size of the energy gap $\Delta$ remains largely unaffected. 

Interestingly our theoretical estimates also predict that the electron spins align ferromagnetically with an effective Heisenberg ferromagnetic interaction $J_F \sim 1$ K (Supplementary section XI). This behavior contrasts with the antiferromagnetic spin alignment typically observed in Mott insulators, where each lattice site is occupied. In our case, the correlated insulator emerges at a fractional filling of the kagome lattice, involving more than one orbital state. Consequently, spin ferromagnetism arises through the Goodenough-Kanamori-Anderson mechanism. These intriguing properties of the observed correlated state, hosted by the electronic kagome lattice, are quite different to states observed in Moir\'{e} systems and open exciting opportunities for future experimental exploration.

In summary, we have established a new platform for creating solid-state artificial crystals, enabling studies of physical phenomena driven by long-range hopping and strong Coulomb interactions. Using this method, we observed a strong insulating state in a kagome flat band, stabilised by long-range Coulomb repulsion and consistent with the model of a loop-current Wigner insulator. The discovery of this unique correlated state is particularly significant, as a true kagome lattice is rarely realized in other artificial solid-state systems, and correlation effects in such systems have yet to be observed in transport experiments. %fThe ability to achieve the distinctive kagome lattice provides a means of investigating a wide array of exotic quantum phenomena, including topological states and various correlated effects, and potentially allowing insights into high-temperature superconductivity~\cite{Yin22}. 

We emphasise that our approach not only allows lattices of any geometry to be created, but is also material agnostic, making it applicable to a variety of 2D systems, including atomically thin materials~\cite{brey_emerging_2009, Zhang21, Yang22, krix_patterned_2023}. Furthermore, the technique can be extended to generate topological systems by introducing spin-orbit interactions through the use of valence band holes instead of conduction band electrons~\cite{Sushkov13}, or extended to study exotic phases in honeycomb and kagome systems~\cite{li_artificial_2020,ghorashi_cano_topological_2023, Yin22} including ferrielectric and topological ferromagnetic states in the high magnetic field regime~\cite{mishra_interactioneffects_Hofstadter2016, mai_interaction-driven_Hofstadter2023}. Overall, the ability to create arbitrary crystal geometries, with unprecedented control over topology, doping, spin-orbit interaction, and superlattice potential opens up the possibility of fabricating and studying an extensive variety of synthetic quantum matter.

\textbf{Acknowledgements}
 We thank G. Khaliullin for helpful discussion. This work was funded by the Australian Research Council Centre of Excellence for Future Low Energy Electronics Technologies (CE170100039) and EP/R029075/1 Non-Ergodic Quantum Manipulation, UK. Device fabrication was partially carried out at the Australian National Fabrication Facility (ANFF) at the UNSW node.

\textbf{Author contributions}
 D.Q.W. fabricated samples and performed transport measurements. D.Q.W., Z.K., O.P.S., A.R.H. and O.K. performed data analysis and discussed the results. A.R.H., O.P.S. and O.K. supervised the project. Z.K., O.P.S., O.A.T. and V.A.T. performed numerical calculations. C.C., I.F. and D.A.R. provided the $\textrm{GaAs/Al}$$_{0.6}$$\textrm{Ga}$$_{0.4}$$\textrm{As}$ heterostructures. D.W., Z.K., O.P.S., A.R.H. and O.K. co-wrote the manuscript with input from all co-authors.

\section{Methods}

{\bf Theoretical methods.} To model the artificial crystal we need to know the shape of the potential experienced by electrons in the 2DEG. A full three-dimensional numerical model of the device in Fig.~\ref{fig:theory}(a), including Hartree screening, has been used in Ref.~\cite{Tkachenko15} to calculate the artificial bandstructure. To simplify the analysis we construct a model Hamiltonian which depends on only a single parameter $W$ which represents the amplitude of the applied potential, yet captures the essential physics and produces a bandstructure in very good agreement with the full 3D numerical solution:

\begin{align}\label{uu}
\begin{split}
        H & = \frac{p^{2}}{2 m^{*}} + U(\bf{r}) \\
    U(\bf{r}) & = 2 W
    \left[
        \cos({\bf G}_1\cdot{\bf r}) + 
        \cos({\bf G}_2\cdot{\bf r}) + 
        \cos({\bf G}_3\cdot{\bf r})
    \right]
\end{split}
\end{align}

Where $\bf{G}_{1,2}$ are the basic reciprocal vectors of the triangular lattice and $\bf{G}_{3} = \bf{G}_{2} - \bf{G}_{1}$ with $|{\bf{G}_{i}}| =  4 \pi / \sqrt{3} a$. Note that the minimum of the potential is $U = - 3 W$ and the maximum is $U = 6 W$, such that the peak-to-peak potential amplitude $U_{p-p}$ is

\begin{align}
    U_{p-p} = 9W
\end{align}

The lattice constant $a$ is 100 nm in all calculations. Here, higher harmonics of the potential have been neglected. The validity of this Hamiltonian has been checked against fully numerical 3D finite element Poisson calculations in Ref.~\cite{Tkachenko15}. It is possible, however, to understand the form of Eqn. \ref{uu} in terms of the following considerations: (i) at the level of the patterned gate the potential has the shape of a triangular array of (circular) hat functions. (ii) According to the Poisson equation, electrons in the plane of the 2DEG experience this potential with Fourier components modified from $U_{k} \to e^{-kz} U_{k}$, where $z$ is the distance to the gate. (iii) The fundamental harmonics have the form $\cos( \bf{G_i} \cdot \bf{r})$, with $|\bf{G}| = 4 \pi / \sqrt{3} a$, and higher harmonics are suppressed relative to these by a factor $e^{-4 \pi / \sqrt{3} a} (\approx 0.1$). (iv) For $W=0$ electrons in the 2DEG are described by a quadratic dispersion, $p^{2} / 2 m^{*}$, where $m^{*} = 0.0667 m_{e}$ is the effective mass of electrons in GaAs. 

% To model an artificial electrostatic crystal we make the following considerations: (i) at the level of the patterned gate the potential has the shape of a triangular array of (circular) hat functions. (ii) According to the Poisson equation, electrons in the plane of the 2DEG experience this potential with Fourier components modified like $U_{k} \to e^{-kz} U_{k}$, where $z$ is the distance to the gate. (iii) The fundamental harmonics have the form, $\cos( \bf{G} \cdot \bf{r})$, with $|\bf{G}| = 4 \pi / \sqrt{3} a$, and higher harmonics are suppressed relative to these by a factor $e^{-4 \pi / \sqrt{3} a} \approx 0.1$. (iv) In the absence of a potential electrons in the 2DEG are described by a quadratic dispersion, $p^{2} / 2 m^{*}$, where $m^{*}$ ($m^{*} = 0.0667 m_{e}$ is the effective mass of electrons in GaAs. Note that the minimum of the potential is $U = - 3 W$ and the maximum is $U = 6 W$, meaning that the total width of the potential is $9W$.

The band structures were computed by exact numerical diagonalisation of the Hamiltonian in Eqn. \ref{uu}. To do this we write Eqn. \ref{uu} in the basis of plane wave states, $| {\bf k} \rangle = e^{i {\bf k} \cdot {\bf r}}$. Since $U({\bf r})$ only mixes states which differ in momentum by $\pm \bf{G}_{i}$
 or $\pm {\bf G}_{2}$ this basis can be restricted to states, $| {\bf k} + {\bf g}_{i} \rangle$, where ${\bf k}$ is a momentum within the first Brillouin zone and ${\bf g}_{i}$ is an arbitrary reciprocal lattice vector. We thus diagonalise the following matrix

\begin{align*}
    \langle {\bf k} + {\bf g}_{i} |
    H
    | {\bf k} + {\bf g}_{j} \rangle
    =
    \frac{({\bf k} + {\bf g}_{i})^{2}}{2 m}
    \delta_{ij}
    +
    W
    \sum_{n = 1}^{3}
    \delta( {\bf g}_{i} - {\bf g}_{j} - {\bf G}_{n})
\end{align*}
Where $\delta_{ij}$ is the Kronecker symbol. This matrix must be truncated such that the $|{\bf g}_{i}|$ are smaller than some upper limit; the limit is chosen such that eigenvalues and eigenvectors are independent of the limit (for example, the limit ${\bf g} = n {\bf G}_{1} + m {\bf G}_{2}$ with $|n|, \ |m| \leq 10$ is more than large enough for the strongest potentials that we consider). Numerical diagonalisation gives a set of energy levels $\varepsilon_{n}({\bf k})$ (these are plotted in Fig. \ref{fig:theory}) and a set of corresponding Bloch functions $\psi_{n,{\bf k}}({\bf r})$. The charge densities shown in Figs. \ref{fig:dirac}e and \ref{fig:kagome}f are equal to $\sum_{\mu_{0} < \varepsilon_{n}({\bf k}) < \mu} |\psi_{n,{\bf k}} ({\bf r})|^{2}$. For Fig. \ref{fig:dirac}e we sum over states from the bottom of band 1 to the mid-point of band 2 ($\mu = 1.5$ meV). For Fig. \ref{fig:kagome}f we sum over states from the bottom of band 3 to the top of band 4, -0.5 meV $ <\epsilon< $ 2.7 meV .
Of course, in the latter case, bands 1 and 2 also contribute to the total charge density, however, these states are not physically relevant when $\mu$ is within the kagome-like bands and we exclude them for the sake of clarity.

\textbf{Experimental methods.} The devices in this study are fabricated on an ultra-shallow high quality undoped $\textrm{GaAs/Al}$$_{0.6}$$\textrm{Ga}$$_{0.4}$$\textrm{As}$ heterostructure (wafer W1740) comprising of a 3 nm GaAs cap, a 22 nm AlGaAs layer, and a thick GaAs buffer layer, grown by molecular beam epitaxy. N-type ohmic contacts (AuGe) to the heterostructure were thermally evaporated into etched pits and then annealed. The triangular lattice is patterned into the surface metal gate by electron-beam lithography and reactive ion etching with $\mathrm{SF_6}$. A 15 nm thick $\mathrm{AlO_x}$ dielectric is deposited with atomic layer deposition to isolate the metal gates from each other and from the ohmic contacts~\cite{Wang20}. The mean free path of electrons in the unpatterned wafer is $3~\mu \mathrm{m}$ at $\mathrm{n=2\times 10^{11}~cm^{-2}}$, much larger than the lattice constant. 
%A measurement of a Hall bar made on the same heterostructure has yielded a 2D electron mobility of 400,000 $\mathrm{cm^2/Vs}$ at $\mathrm{n=2.5\times 10^{11} cm^{-2}}$ and $\mathrm{T= 250 mK}$. 
Transport measurements were performed with standard lock-in techniques between 17 to 33 Hz. All devices are cooled with all gates grounded. Because of the undoped heterostructure used, these devices are very stable between different cool downs.  Measurements are done on two types of device designs: a Van der Pauw geometry with a patterned area of $5 \times 5~\mu \mathrm{m}$ and a Hallbar geometry with a $2~\mu \mathrm{m}$ wide channel.

%% file: supplementary.tex
\title{Supplementary Information for Artificial electrostatic crystals: a new platform for creating correlated quantum states}

\author{Daisy Q. Wang} 
\affiliation{School of Physics, The University of New South Wales, Sydney, NSW 2052, Australia}
\affiliation{Australian Research Council Centre of Excellence in Future Low-Energy Electronics Technologies,The University of New South Wales, Sydney, NSW 2052, Australia}

\author{Zeb Krix}
\affiliation{School of Physics, The University of New South Wales, Sydney, NSW 2052, Australia}
\affiliation{Australian Research Council Centre of Excellence in Future Low-Energy Electronics Technologies,The University of New South Wales, Sydney, NSW 2052, Australia}

\author{Olga A. Tkachenko}
\affiliation{Rzhanov Institute of Semiconductor Physics, Novosibirsk, 630090, Russia}

\author{Vitaly A. Tkachenko}
\affiliation{Rzhanov Institute of Semiconductor Physics, Novosibirsk, 630090, Russia}
\affiliation{Novosibirsk State University, Novosibirsk, 630090, Russia}

\author{Chong Chen}
\affiliation{Cavendish Laboratory, J. J. Thomson Avenue, Cambridge, CB3 0HE, United Kingdom}

\author{Ian Farrer\footnote{Present Address: Department of Electronic and Electrical Engineering, The University of Sheffield, Mappin Street, Sheffield, S1 3JD, United Kingdom.} }  
\affiliation{Cavendish Laboratory, J. J. Thomson Avenue, Cambridge, CB3 0HE, United Kingdom}

\author{David A. Ritchie}
\affiliation{Cavendish Laboratory, J. J. Thomson Avenue, Cambridge, CB3 0HE, United Kingdom}

\author{Oleg P. Sushkov}
\affiliation{School of Physics, The University of New South Wales, Sydney, NSW 2052, Australia}
\affiliation{Australian Research Council Centre of Excellence in Future Low-Energy Electronics Technologies,The University of New South Wales, Sydney, NSW 2052, Australia}

\author{Alexander R. Hamilton}
\affiliation{School of Physics, The University of New South Wales, Sydney, NSW 2052, Australia}
\affiliation{Australian Research Council Centre of Excellence in Future Low-Energy Electronics Technologies,The University of New South Wales, Sydney, NSW 2052, Australia}

\author{Oleh Klochan}
\affiliation{School of Physics, The University of New South Wales, Sydney, NSW 2052, Australia}
\affiliation{Australian Research Council Centre of Excellence in Future Low-Energy Electronics Technologies,The University of New South Wales, Sydney, NSW 2052, Australia} 
\affiliation{School of Science, The University of New South Wales, Canberra, ACT 2612, Australia}

\date{\today}
\maketitle

\renewcommand{\thefigure}{S\arabic{figure}}
\setcounter{figure}{0}    

\section{Comparison of our device with previous studies}
To make the artificial crystal and observe band structure effects it is critical to create a highly uniform periodic potential with an amplitude larger than the Fermi energy while keeping the disorder broadening of the energy levels low. To highlight  the advantages of our device architecture we compare the carrier mobility after patterning vs the artificial lattice constant for systems where the modulation potential is strong in Fig. S1(a), including early works in etched GaAs system and etched graphene systems. Etching is known to produce a strong modulation potential but at the same introduces disorder that causes unwanted scattering and limits the carrier mobility. This makes it impossible to observe the artificial bandstructure. In our case, the sample mobility of $\sim500,000$ cm$^{2}$/Vs is measured in the patterned device at a high density $2\times10^{11}$ cm$^{-2}$ where superlattice effects are negligible. The same measurement gives $\sim300,000$ cm$^{2}$/Vs at $1.5\times10^{11}$ cm$^{-2}$. As a comparison devices with an unpatterned 2D top gate yield a similar mobility at those densities. This confirms that the fabrication process has very little damage to the heterostructure.  

\begin{figure}[h!]
\centering
\includegraphics[width=\textwidth]{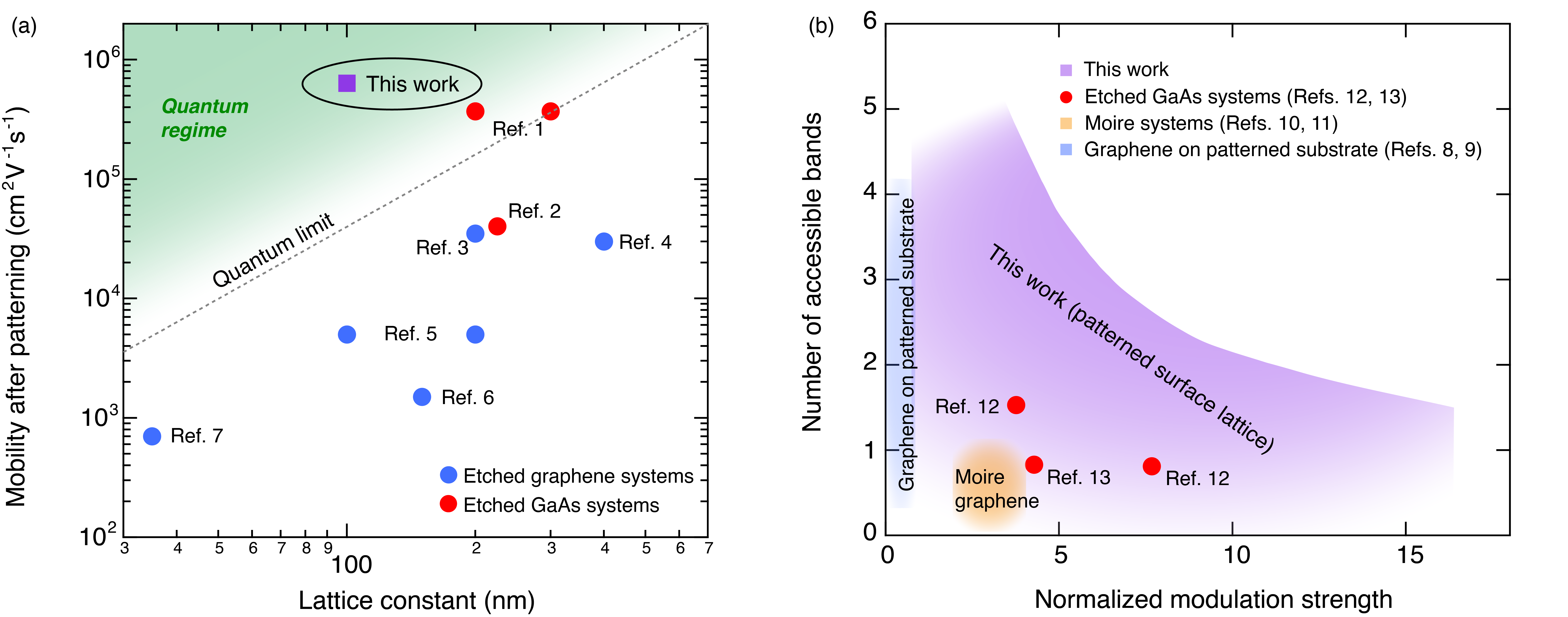}
\caption{
(a) Carrier mobility of the patterned system versus artificial lattice constant for etched systems with strong modulation potential. Red circles -  etched GaAs heterostructures (Refs.~\onlinecite{ Weiss91_APL, Simoni10}), blue circles - etched graphene systems (Refs.~\onlinecite{Sandner15, Yagi15, Peters13, Pan17, Jessen19}), violet square - our work. The green shaded region denotes the quantum regime, where the mean free path is bigger than the artificial lattice constant and new bandstructure effects can be observed. The dashed line marks the boundary between the classical and quantum regimes where the two length scales are equal.
(b) The number of accessible mini-bands versus the normalized modulation potential strength $U_{pp}/E_K$. Shaded blue area - graphene on a patterned substrate (Refs.~\onlinecite{Forsythe18, Huber22}), shaded orange area - graphene Moire systems (Refs.~\onlinecite{Yankowitz12, Cao18}), red dots - etched GaAs heterostructures (Refs.~\onlinecite{Wang_Pinczuk2018, Du21}). The shaded violet area indicates the regime acccessible in the present work, which provides a system in which artifical bands are clearly resolved (the potential modulation is much larger than the kinetic energy) while also allowing the Fermi energy to be tuned through multiple artificial bands.}
\label{fig:comp}
\end{figure} 

To highlight the tunability of our device architecture we compare our system with other systems where artificial bands have been observed previously in Fig. S1(b). We plot the number of accessible bands against the normalised strength of the modulation potential, defined as the peak-to-peak modulation potential $U_{pp}$ divided by the characteristic kinetic energy $E_{K}$. Note that $E_K= g^2/2m$ for GaAs and $E_K= v_F g$ for graphene where $g$ is the reciprocal lattice vector of the artificial crystal. Our system allows access to an entirely new regime, where the modulation is strong $U_{pp} \gg E_K$ yet the Fermi energy can be continuously tuned through multiple bands.

\section{Measurements in the weak modulation regime}

\begin{figure}[h!]
\centering
\includegraphics[width=0.85\textwidth]{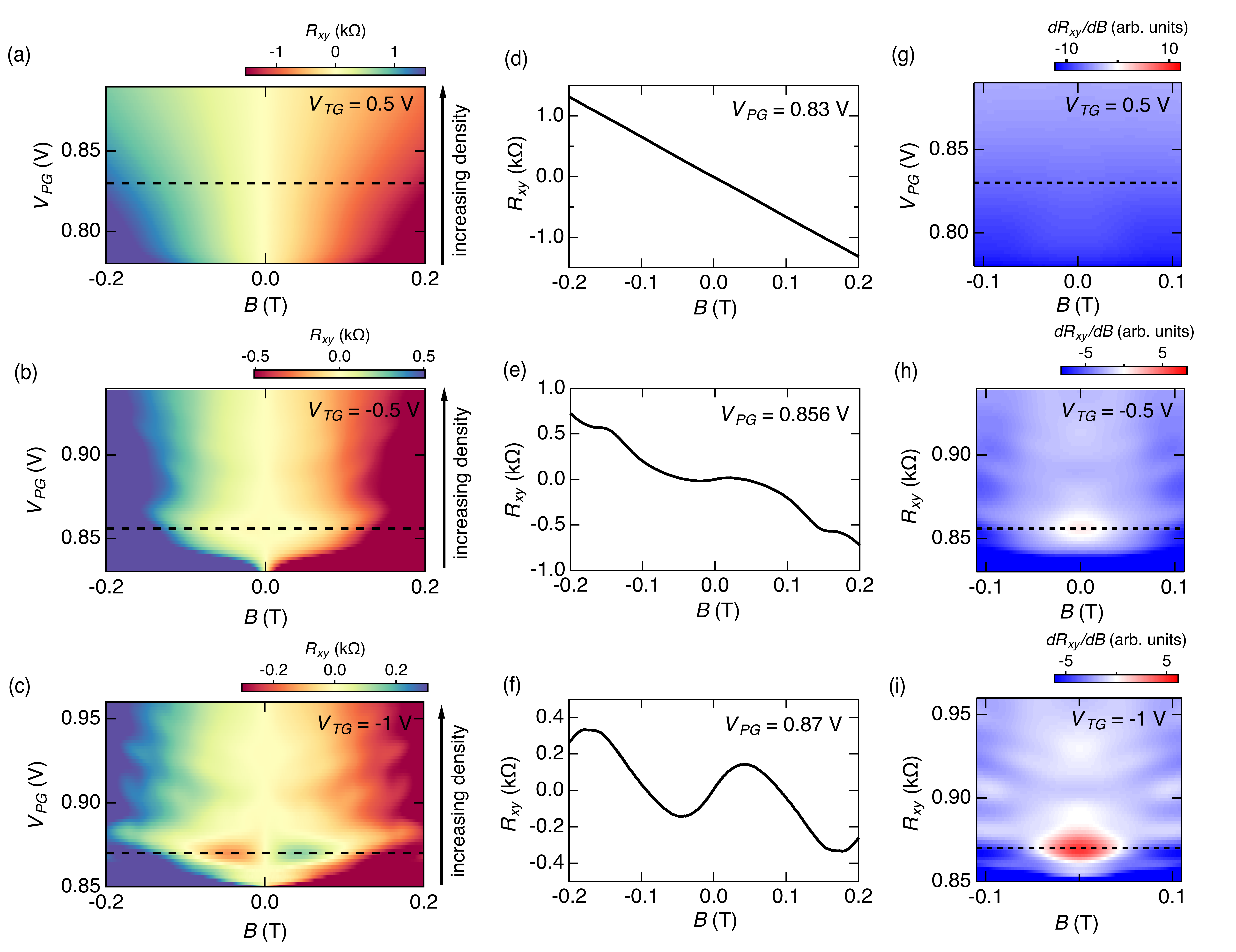}
\caption{Hall measurements of device D37 on the 37 nm deep heterostructure W916 with a lattice constant of 120 nm at $T=1.5$ K. First column shows the evolution of Hall resistance as a function of the patterned gate voltage $V_{PG}$ and perpendicular magnetic field B with increasing modulation strength (a) $V_{TG}=0.5~\mathrm{V}$, (b) $V_{TG}=-0.5~\mathrm{V}$ and (c) $V_{TG}=-1~\mathrm{V}$. Dashed lines indicate the position of the linecuts for the second column. (d)-(f) Hall resistance as a function of B field at (d) $V_{PG}=0.83~\mathrm{V}$, (e) $V_{PG}=0.858~\mathrm{V}$ and (f) $V_{PG}=0.87~\mathrm{V}$ for three modulation strengths respectively, showing the change in Hall resistance near zero-field as the modulation strength is increased. (g)-(i) Hall coefficient $R_H=dR_{xy}/dB$ as a function of as a function of $V_{PG}$ and B at three different modulation strengths corresponding to (a)-(c). Red regions indicate where the Hall slope changes sign and the carrier type becomes hole-like.}
\label{fig:Hall}
\end{figure}

To access the very weak modulation regime, it is necessary to reduce the superlattice potential strength by using a deeper heterostructure. The deeper the heterostructure is, the weaker the superlattice potential. Details about the effect of the wafer depth on the strength of the superlattice potential are discussed in Supplementary section III. Here we show measurements performed on a relatively deep device fabricated using a 37 nm deep heterostructure (W916) and show how the device can be continuously tuned from a plain 2DEG to an artificial crystal. 

In the regime of very weak modulation strength $V_{TG}=+0.5~\mathrm{V}$, the system bears resemblance to a conventional 2D electron gas, as demonstrated by the featureless Hall resistance in Fig.~\ref{fig:Hall}(a),(d) and a negative Hall coefficient in Fig.~\ref{fig:Hall}(g). This slope is inversely proportional to the density of the 2D electron gas (2DEG) $n=1/(edR_{xy}/dB)$, which in turn is directly proportional to the applied voltage on the patterned gate $V_{PG}$.

When the modulation potential is increased, by setting a top-gate voltage to $V_{TG}=-0.5~\mathrm{V}$, we observe suppression of the Hall resistance around $B=0$ (Fig.~\ref{fig:Hall}(b)). Increasing the potential modulation further to $V_{TG}=-1~\mathrm{V}$ causes the Hall resistance around $B=0$ to change sign, as shown in Fig.~\ref{fig:Hall}(c,f). As a result, the Hall coefficient switches sign as highlighed by the red-colored region in Fig.~\ref{fig:Hall}(i), a direct signature of the existence of hole-like carriers. 
%The periodic oscillations of the Hall resistance as a function of the magnetic field $B$ seen in Fig.~\ref{fig:Hall}(f) are due to gauge invariance when integer numbers of flux quanta are threaded through each artificial unit cell, as discussed in the main text.

\section{Bandstructure evolution and Calibration of the filling factor}

\begin{figure}[h!]
\centering
\includegraphics[width=1.1\textwidth]{W1740_transition_v2.png}
\caption{Measured Hall Coefficient $R_H=dR_{xy}/dB$ as a function of patterned gate voltage $V_{PG}$ and perpendicular magnetic field $B_{\perp}$ for device D251 on 25 nm deep heterostructure W1740 with a lattice constant of 100 nm at $T=1.5$ K. The modulation strength is gradually increased by stepping up the top-gate voltage $V_{TG}$: (a) $V_{TG}=0~\mathrm{V}$, (b) $V_{TG}=-0.5~\mathrm{V}$, (c) $V_{TG}=-1~\mathrm{V}$, (d) $V_{TG}=-1.5~\mathrm{V}$, (e) $V_{TG}=-2~\mathrm{V}$, (f) $V_{TG}=-2.5~\mathrm{V}$ and (g) $V_{TG}=-3~\mathrm{V}$. Black dashed lines indicate the positions of the Dirac points (DP1 and DP2) and van Hove singularities (VH1, VH2, VH3) where Hall slope changes sign. Red dashed lines indicate the positions where each band terminates. Since density is linearly proportional to $V_{PG}$, the red dashed lines are evenly spaced with a band size of $\Delta V_{PG}\approx 24$ meV.}
\label{fig:evolution}
\end{figure} 

%\begin{figure}[h!]
%\centering
%\includegraphics[width=\textwidth]{W1740_transition.png}
%\caption{Measured Hall Coefficient $R_H=dR_{xy}/dB$ as a function of patterned gate voltage $V_{PG}$ and perpendicular magnetic field $B_{\perp}$ for device D251 on 25 nm deep heterostructure W1740 with a lattice constant of 100 nm at $T=1.5$ K. The modulation strength is gradually increased by stepping up the top-gate voltage $V_{TG}$. (a) $V_{TG}=0~\mathrm{V}$, (b) $V_{TG}=-0.5~\mathrm{V}$, (c) $V_{TG}=-1~\mathrm{V}$ and (d) $V_{TG}=-1.5~\mathrm{V}$ in the first row focus on the graphene bands at low densities (low $V_{PG}$) with transitions VH1, DP1 and VH2. (e) $V_{TG}=-1.5~\mathrm{V}$, (f) $V_{TG}=-2~\mathrm{V}$, (g) $V_{TG}=-2.5~\mathrm{V}$ and (h) $V_{TG}=-3~\mathrm{V}$ in the second row show the development of the kagome bands (VH3 and DP2). Black dashed lines indicate the positions of the Dirac points and van Hove singularities where Hall slope changes sign. Red dashed lines indicate the positions where each band terminates. Since density is linearly proportional to $V_{PG}$, the red dashed lines are evenly spaced with a band size of $\Delta V_{PG}\approx 24$ meV.}
%\label{fig:evolution}
%\end{figure} 

In Fig.~\ref{fig:evolution} we show the measured Hall coefficient $R_H$ for different modulation strengths. Band filling can be determined by tracing the positions of the van Hove singularities(VH) and Dirac points (DP) where $R_H$ changes sign. As the modulation strength gradually increases by applying a more negative gate voltage on the top-gate, the mini-bands in the system gradually move apart and become separate, which leads to more sign changes at high densities (high $V_{PG}$). The size of one band $\Delta V_{PG}$ can be extracted either from DP1 to the end of the hole-like (red) region at the end of the second band which corresponds to filling one band (Fig.~\ref{fig:evolution}(d)), or from the end of this hole-like region to the second Dirac point (DP2) which corresponds to filling two band (Fig.~\ref{fig:evolution}(f)-(g)). Using these methods, we extract the band density $n_0$ of the system in $V_{PG}$ is $\Delta V_{PG}\approx 24~\mathrm{mV}$, which for a $a=100$ nm triangular lattice, is equivalent to a 2DEG density of $2.3\times10^{10}~\mathrm{cm^{-2}}$ with filling factor $\nu = 2$ electrons per band from spin. This provides a simply conversion between $\Delta_{PG}$ to electron filling factor $\nu$, i.e. 2 electrons per unit cell for every $24$ mV on $V_{PG}$. This conversion value of $V_{PG}$ is only related to the wafer depth (when lattice constant is kept constant) as the device can be modelled by a simple parallel plate capacitor (supplementary section~\ref{sec:cap}). 

There are two subtleties to pay attention to for the band assignment. First, at an intermediate modulation strength (Fig.~\ref{fig:evolution}(b)-(c)) the end of the second band can overlap with the third band (Fig. 1(d)), in which case a negative Hall coefficient can still occur in the presence of both electron-like and hole-like carriers. This is the reason why the positive Hall slope region above the second van Hove singularity can span over a larger $\Delta V_{PG}$ than expected. When the second band is fully separate from the third band, the size of the positive Hall coefficient region above the first Dirac point can be used to calibrate the gate voltage for one band density. Secondly, as the modulation strength increases (Fig.~\ref{fig:evolution}(e)-(g)), we gradually lose transport in the first band and the sign changes in the first two graphene-like bands become suppressed. This is possibly due to disorder in modulation potential and localization of electrons in the strong modulation regime. However, from tracing the evolution, we can still identify the features in Hall resistance corresponding to the hole-like region at the end of the second band, which provides the first anchor point in the strong modulation regime.

In Fig.~\ref{fig:evolution2} we show the evolution of the bandstructure of another device D33 fabricated on a different wafer which is a 33nm deep heterostructure with a lattice constant 100 nm. Behaviour of the device is very similar to D251 except it shows a slightly weaker modulation strength overall. The strength of the modulation potential is predominantly determined by two experimental factors that are related to the device architecture and design: 1) the distance between the patterned gate and the 2DEG $d$ and 2) the lattice constant $a$ (see Ref.~\onlinecite{Tkachenko15}). When $d$ is reduced, the modulation strength increases exponentially. Increasing $a$ also increases potential modulation. However, a larger lattice constant results in a smaller band density which hinders the observation of the mini-bands. Therefore, the optimal range for the lattice constant is $\sim100$ nm with a band density of $\sim2\times10^{10}~\mathrm{cm^{-2}}$ which is easily accessible in GaAs/AlGaAs heterostructures. 

A larger $d$ also decreases the capacitance of PG, which for the same lattice constant and band density, proportionally increases the band size $\Delta V_{PG}$. From the 33nm deep heterostructure we extract a band filling of $\Delta V_{PG} \approx 32$ mV. The change in $\Delta V_{PG}$ per band compared to the 25nm deep heterostructure (Fig.~\ref{fig:evolution}, $\Delta V_{PG}=24$ mV) exactly reflects the change in depth assuming a simple parallel capacitor $\Delta V_{PG}/d \propto n_{0} = 2.3\times10^{10} ~cm^{-2}$.

\begin{figure}[h!]
\centering
\includegraphics[width=1.1\textwidth]{W1735_transition_v2.png}
\caption{Hall Coefficient $R_H=dR_{xy}/dB$ as a function of patterned gate voltage $V_{PG}$ and perpendicular magnetic field $B_{\perp}$ for device D33 on 33 nm deep heterostructure W1735 with a lattice constant of 100 nm at $T=1.5$ K. The modulation strength is gradually increased by stepping up the top-gate voltage $V_{TG}$: (a) $V_{TG}=0~\mathrm{V}$, (b) $V_{TG}=-0.5~\mathrm{V}$, (c) $V_{TG}=-1~\mathrm{V}$, (d) $V_{TG}=-1.5~\mathrm{V}$, (e) $V_{TG}=-2~\mathrm{V}$, (f) $V_{TG}=-2.5~\mathrm{V}$ and (g) $V_{TG}=-3~\mathrm{V}$. Black dashed lines indicate the positions of the Dirac points (DP1 and DP2) and van Hove singularities (Vh1, VH2, VH3) where Hall slope changes sign. Red dashed lines indicate the positions where each band terminates. Since density is linearly proportional to $V_{PG}$, the red dashed lines are evenly spaced with a band size of $\Delta V_{PG}\approx 32$ meV.}
\label{fig:evolution2}
\end{figure} 

\section{Reproducibility across multiple devices}

We have measured multiple devices on three different heterostructures: W1740 which is 25 nm deep (Figs in the main text and Fig.S3, S5, S6, S7), W1735 which is 33 nm deep (Fig.S4) and W916 which is 37 nm deep (Fig.S2) as detailed in Table~\ref{table:device}. The devices on 25 nm and 33 nm deep wafers were made with a 100 nm lattice while the device on the 37 nm deep heterostructure has a 120 nm lattice. Even though the accessible range of modulation strength is different for different wafers, we observe very similar behaviour across all devices. 

\begin{table}[h!]
\centering
\begin{tabular}{ |p{3cm}|p{3cm}|p{3cm}|p{3cm}|p{3cm}| }
 \hline
Device Name & Wafer Number & Wafer Depth & Lattice Constant & Device Geometry \\
 \hline
 D251   & W1740    & 25 nm &   100 nm & Van der Pauw\\
 D252 &   W1740  & 25 nm   & 100 nm & Hallbar\\
 D253 & W1740 & 25 nm &  100 nm & Hallbar \\
 D254    & W1740  & 25 nm &  100 nm & Hallbar \\
 D37 &   W916  & 37 nm & 120 nm & Van der Pauw\\
 D33 & W1735  & 33 nm   & 100 nm & Van der Pauw \\
 \hline
\end{tabular}
    \caption{List of devices.}
    \label{table:device}
\end{table}

\begin{figure}[h!]
\centering
\includegraphics[width=0.85\textwidth]{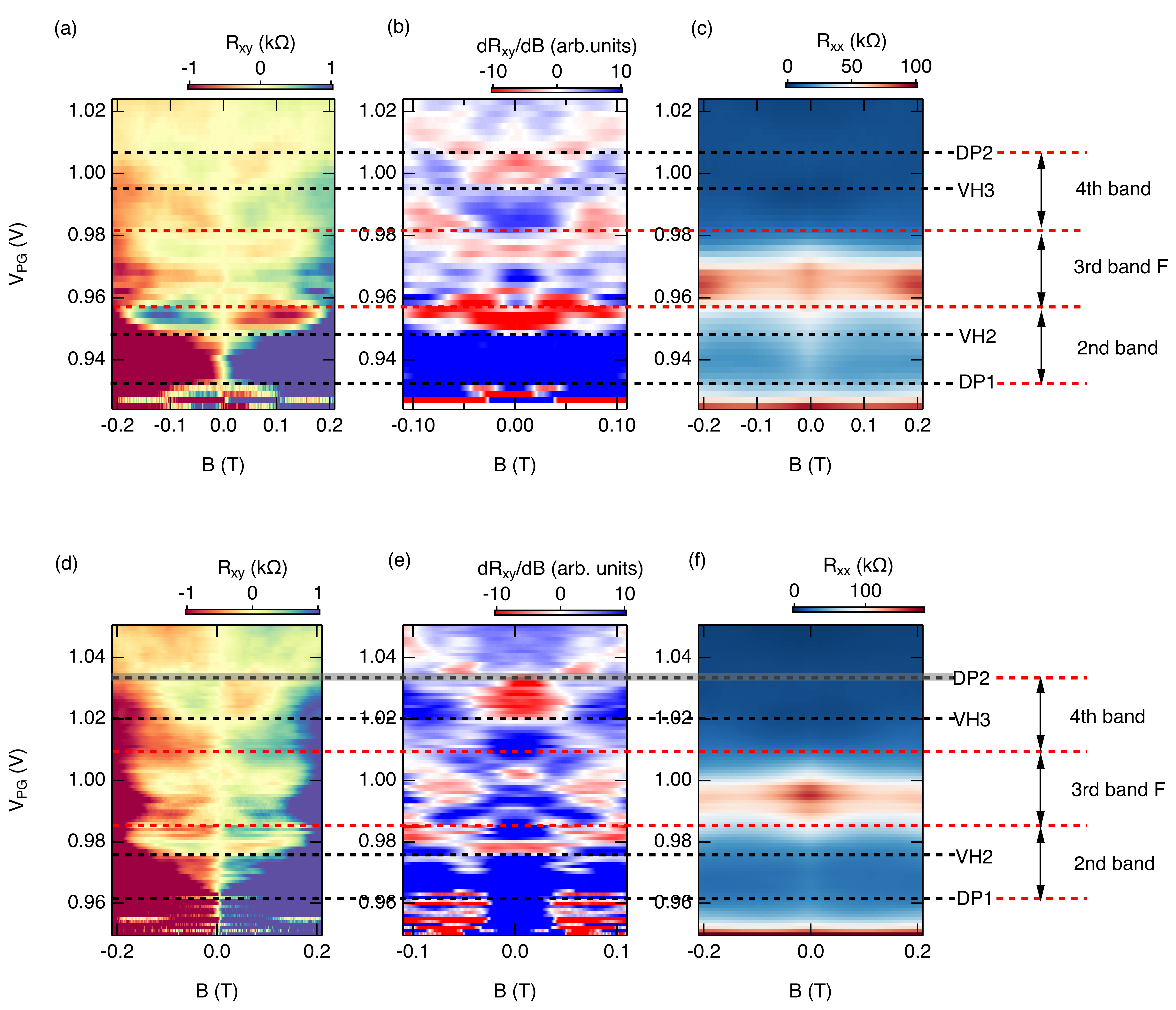}
\caption{Hall resistance $R_{xy}$, Hall coefficient $R_H$ and longitudinal resistance $R_{xx}$ as a function of patterned gate voltage $V_{PG}$ and perpendicular magnetic field $B_{\perp}$ for device D252 on 25 nm deep heterostructure W1740 with a lattice constant of 100 nm at $T=1.5$ K. Modulation strength is kept at $V_{TG}=-1.5$ V (a-c) and $V_{TG}=-2$ V (d-f). In both cases the kagome bands are developed. Black dashed lines indicate the positions of the Dirac points and the van Hove singularities where Hall slope changes sign. Red dashed lines indicates the band filling. Grey shaded area in (d-f) indicates the uncertainty $\sim 8\%$ $(2~\mathrm{mV})$ of band assignment in determining position of anchor point DP2.}
\label{fig:D252}
\end{figure} 

\begin{figure}[h!]
\centering
\includegraphics[width=0.85\textwidth]{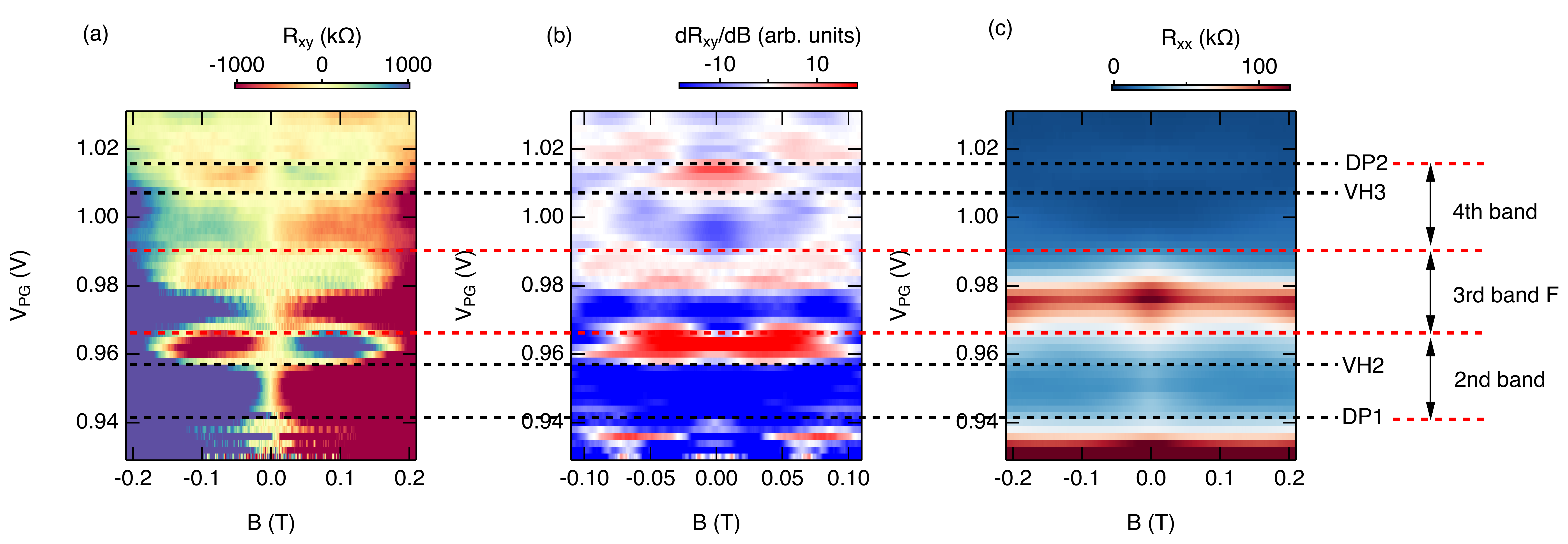}
\caption{Hall resistance $R_{xy}$ (a), Hall coefficient $R_H$ (b) and longitudinal resistance $R_{xx}$ (c) as a function of patterned gate voltage $V_{PG}$ and perpendicular magnetic field $B_{\perp}$ for device D253 on 25 nm deep heterostructure W1740 with a lattice constant of 100 nm at $T=1.5$ K. Modulation strength is kept at $V_{TG}=-1.5$ V where the kagome bands are developed. Black dashed lines indicate the positions of the Dirac points and the van Hove singularities where Hall slope changes sign. Red dashed lines indicates the band filling.}
\label{fig:D253}
\end{figure} 

\begin{figure}[h!]
\centering
\includegraphics[width=0.85\textwidth]{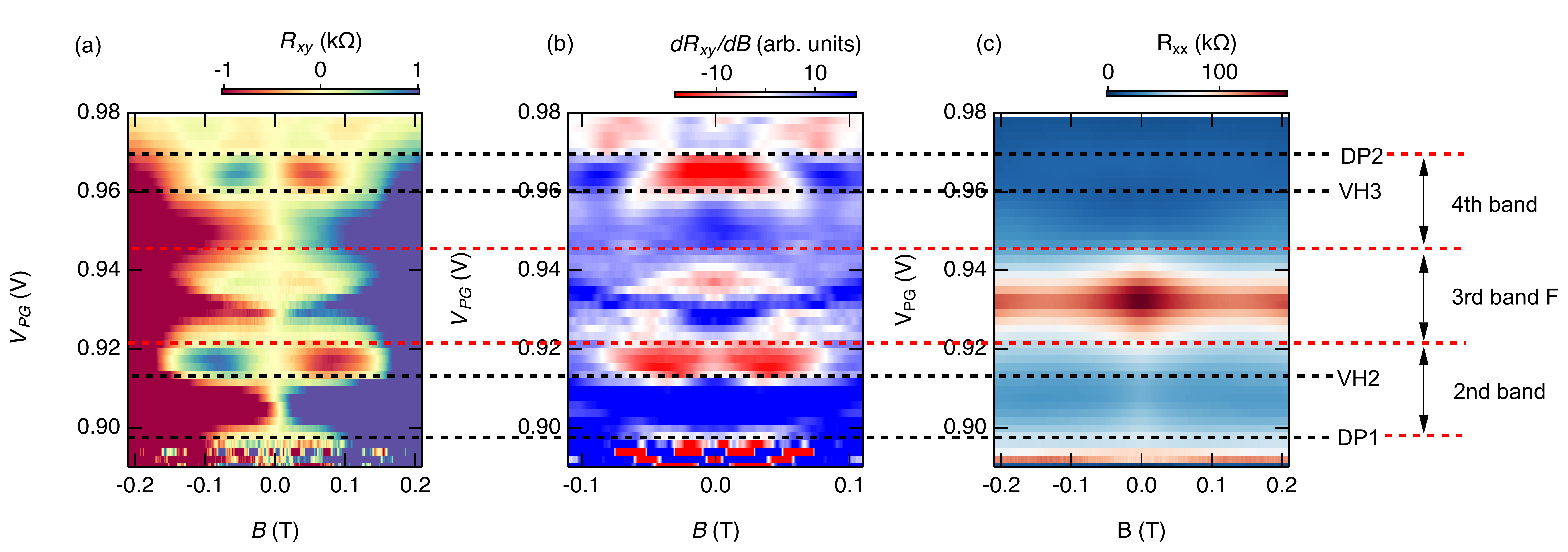}
\caption{Hall resistance $R_{xy}$ (a), Hall coefficient $R_H$ (b) and longitudinal resistance $R_{xx}$ (c) as a function of patterned gate voltage $V_{PG}$ and perpendicular magnetic field $B_{\perp}$ for device D254 on 25 nm deep heterostructure W1740 with a lattice constant of 100 nm at $T=1.5$ K. Modulation strength is kept at $V_{TG}=-1.5$ V where the kagome bands are developed. Black dashed lines indicate the positions of the Dirac points and the van Hove singularities where Hall slope changes sign. Red dashed lines indicates the band filling.}
\label{fig:D254}
\end{figure} 

On the shallowest heterostructure (W1740, 25 nm deep) where modulation strength is the strongest we have measured multiple devices D251 (Van der Pauw geometry), D252 (Hall bar geometry, 2 $\mu$m wide channel), D253 (Hall bar geometry, 2 $\mu$m wide channel) and D254 (Hall bar geometry, 3 $\mu$m wide channel). In Fig.~\ref{fig:D252}(a-c), Fig~\ref{fig:D253} and Fig.~\ref{fig:D254} we show the measured Hall resistance $R_{xy}$, Hall coefficient $R_H=dR_{xy}/dB$ and longitudinal resistance $R_{xx}$ at $V_{TG}=-1.5$ V where the band assignment is performed. All devices show the same band size $\Delta V_{PG} \approx 24$ mV as expected and a resistance peak at the centre of the flat band. In Fig.~\ref{fig:D252}(d-f), we also show the measurement done at a stronger modulation $V_{TG} = -2$ V, which is the configuration used in Fig. 4 of the main text. Because of the effects of the strong modulation on the Hall signal at low densities, the band assignment here relies on the anchor point DP2. Once the position of DP2 is determined, the position of the flat band is known as the band size is $24$ mV. Here we estimate the error in determining the position of DP2 is $\lesssim 2$ mV as highlighted by the grey shaded area in Fig.~\ref{fig:D252}(d-f), which is $8\%$ of the band density. 

\section{Behaviour of the band gap}\begin{figure}[h!]

\centering
\includegraphics[width=\textwidth]{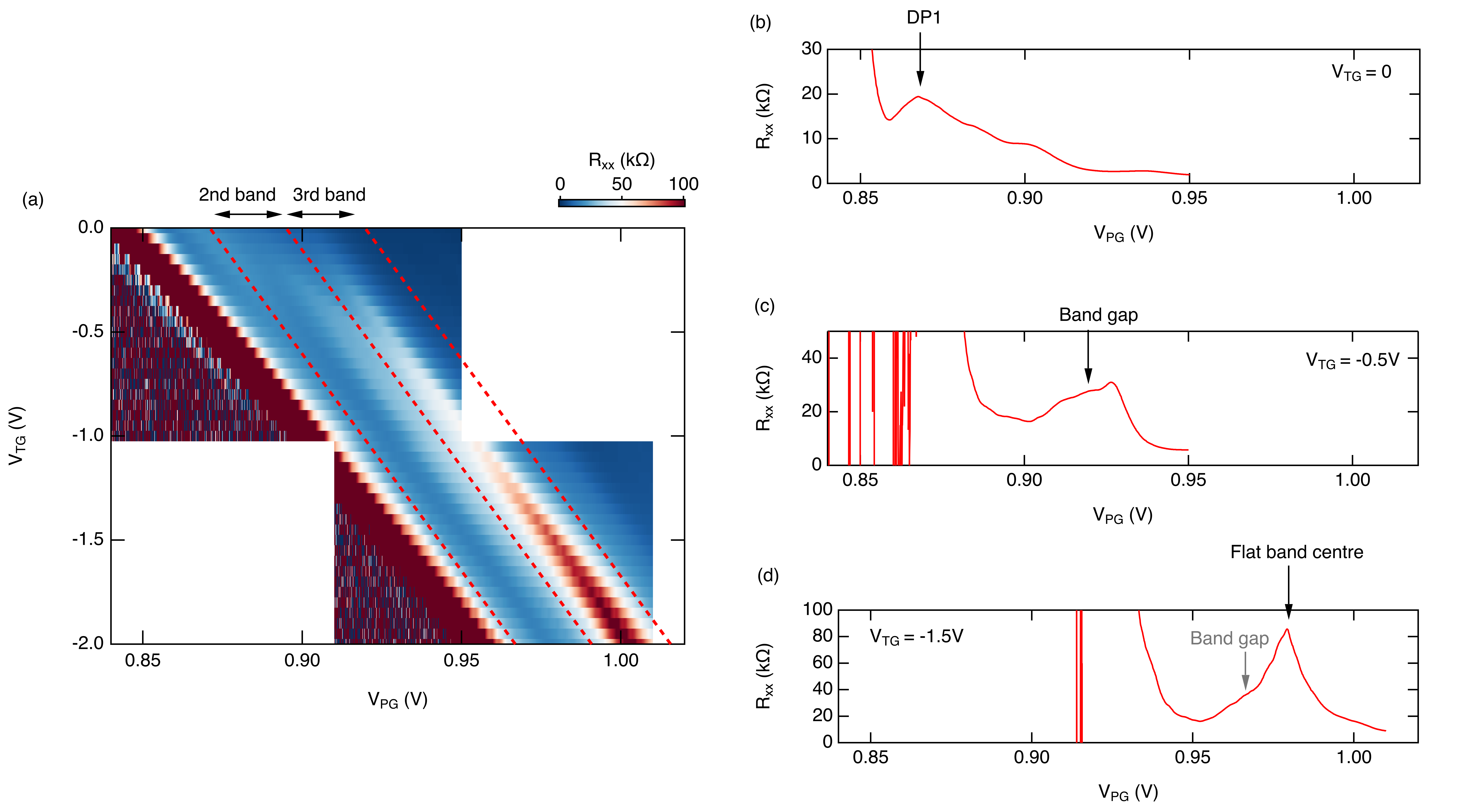}
\caption{(a) Longitudinal resistance $R_{xx}$ as a function of patterned gate voltage $V_{PG}$ ($\propto$ density) and modulations strength $V_{TG}$ for device D253 at $T=1.5$ K. Red dashed lines indicate the band filling. (b-d) Line cuts of $R_{xx}$ at $V_{TG} = 0$, -0.5 V and -1.5 V respectively with black arrows indicating the strongest resistance peak position. The grey arrow in (d) indicates the position of the band gap for the strong modulation case.}
\label{fig:D253_TG}
\end{figure} 

In Fig.~\ref{fig:D253_TG} we show the longitudinal resistance of Device D253 while continuously tuning the modulation strength. As $V_{TG}$ becomes more negative (increasing modulation), the band filling corresponding to the same density shifts to higher $V_{PG}$ due to simple capacitive coupling between the two gates. At different modulation strengths, the strongest $R_{xx}$ peak in the system is generated by different mechanisms. At $V_{TG} = 0$ V (Fig.~\ref{fig:D253_TG}(b)), the resistance peak at the first Dirac point (DP1) appears most resistive. Increasing $V_{TG}$ to $-0.5$ V sees a broad peak developing near the band gap between the second and third bands (Fig.~\ref{fig:D253_TG}(c)). Finally, at $V_{TG} = -1.5$ V, the strong resistance peak at the flat band centre dominates with a resistance much larger than either the Dirac peak or the band gap peak (Fig.~\ref{fig:D253_TG}(d)). A hint of resistance rise can still be seen in Fig.~\ref{fig:D253_TG}(d) where the position of the band gap is indicated by the grey arrow. However, because of the strong resistance peak at the flat band centre, features of the band gap becomes less visible when the kagome flat band becomes fully developed.

\section{Measurements of different configurations on device D252}

\begin{figure}[h!]
\centering
\includegraphics[width=\textwidth]{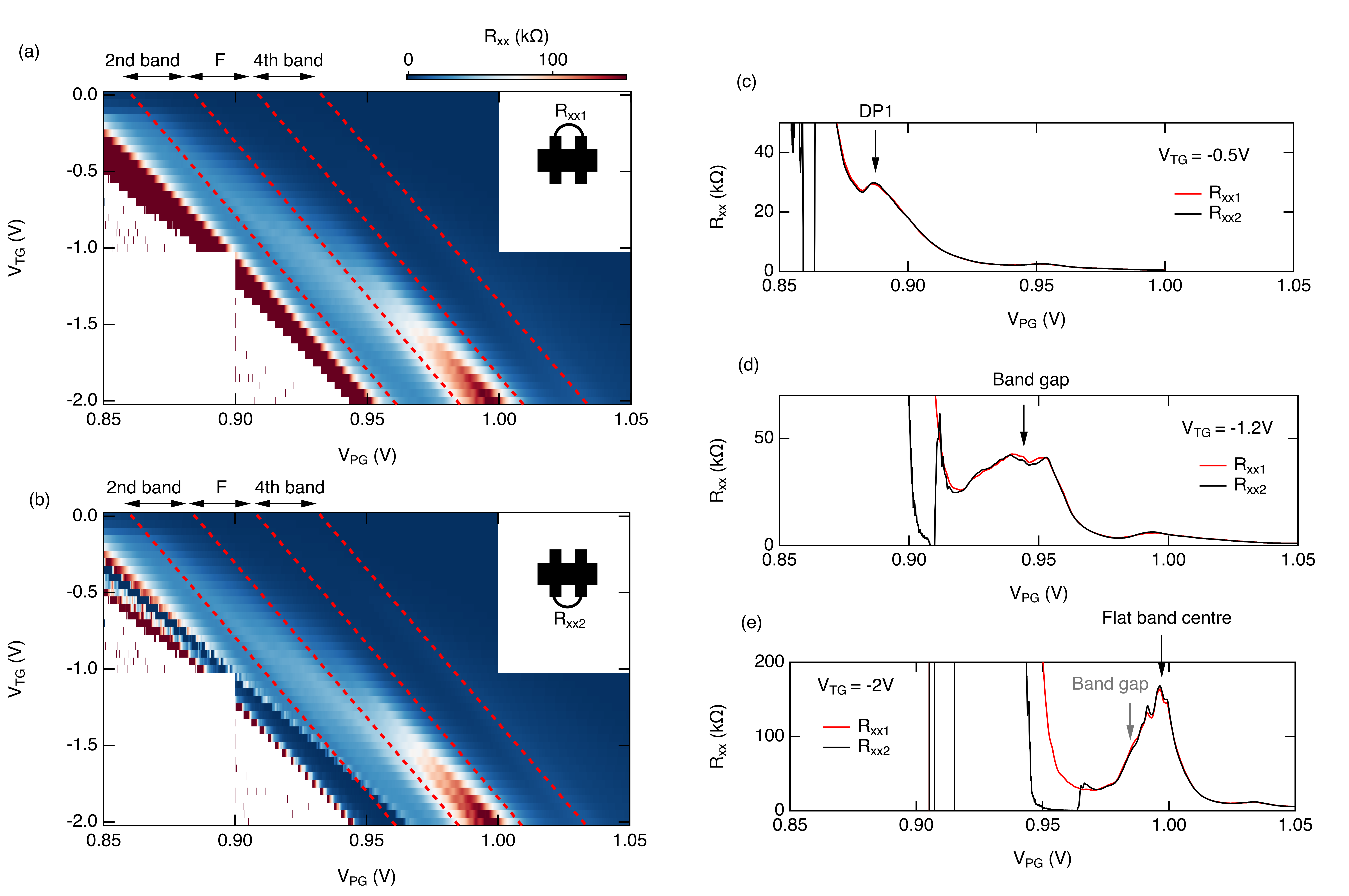}
\caption{Longitudinal resistance $R_{xx}$ as a function of patterned gate voltage $V_{PG}$ ($\propto$ density) and modulations strength $V_{TG}$ for device D252 at $T=1.5$ K in two configurations $R_{xx1}$ (a) and $R_{xx2}$ (b) as illustrated in the insets. In both configurartions, the distance between the contact leads is $2 \mu$m. Red dashed lines indicate the band filling. (c-e) Line cuts of $R_{xx}$ at $V_{TG} = -0.5$, -1.2 V and -2 V respectively with black arrows indicating the strongest resistance peak position. The grey arrow in (d) indicates the position of the band gap for the strong modulation case.} 
\label{fig:D252_TG}
\end{figure}

In this section we show the results from a second $R_{xx}$ configuration on device D252 as a supplement to the results shown in the main text. As shown by the schematics in Fig.~\ref{fig:D252_TG}, $R_{xx}$ can be measured using two different pairs of contacts on two sides of the Hall bar. In the main text (Fig.4), $R_{xx1}$ is presented. In Fig.~\ref{fig:D252_TG} we show both $R_{xx1}$ and $R_{xx2}$ while continuously tuning the modulation strength via $V_{TG}$. The same behaviour as Device D253 (Section V) is observed in both pairs. The only discrepancy between the two configurations is that contacts used in $R_{xx2}$ stop working at low densities earlier than the ones used in $R_{xx1}$. This is signaled by a zero resistance in $R_{xx2}$ near pinch-off.

\begin{figure}[h!]
\centering
\includegraphics[width=\textwidth]{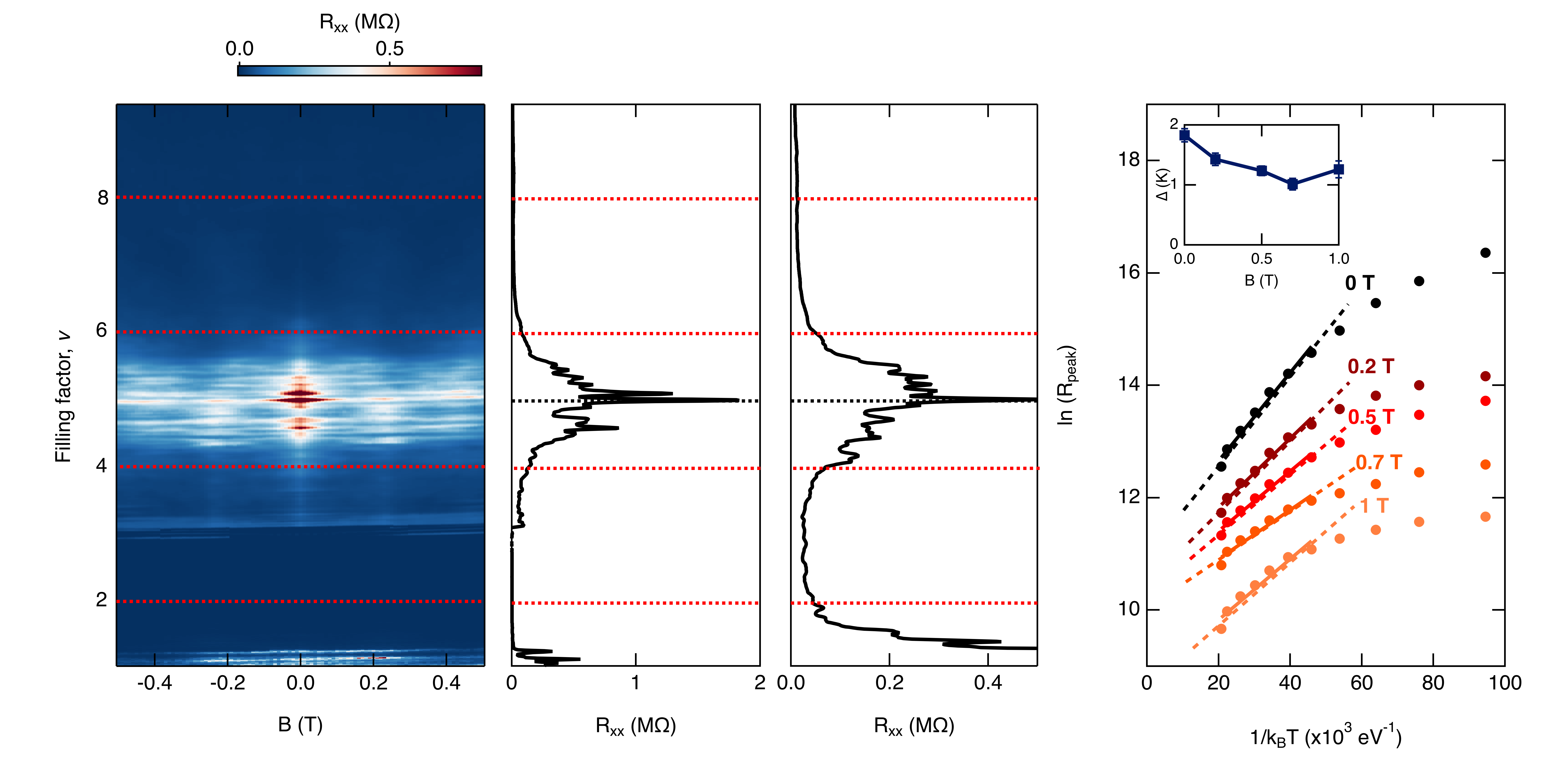}
\caption{ (a) Measured longitudinal resistance of device D252 with $V_{PG}=-2$ V at $T=350$ mK in configuration $R_{xx2}$. Red dashed lines indicate the band filling. Line cuts of $R_{xx2}$ at different magnetic field $B_{\perp}=0$ (b) and $B_{\perp}=0.1$ T (c) with black dashed lines indicate half filling of the flat band. (c) Arrhenius plot (circles) of the resistance of the half-filling insulating state at different $B_{\perp}$. The solid lines (with dashed extrapolation) show the fitting with $\exp{[ \,-\Delta/(2k_BT)] \,}$. The inset shows the extracted thermal activation gap $\Delta$ as a function of $B_{\perp}$. }
\label{fig:D252_peak2}
\end{figure}

In Fig.~\ref{fig:D252_peak2} we show the same measurement as in Fig.4 for configuration $R_{xx2}$. Similarly, the zero resistance at low densities ($\nu < 3$) is attributed to ohmic contacts stop working. However, this should not affect the behavior of the device in the flat band where we are interested in. A strong resistance peak at the half filling of the flat band is also observed in $R_{xx2}$ with a thermal activation gap of $1 - 2$ K similar to $R_{xx1}$.

\section{Coulomb screening, effect on modulation strength}

The actual potential ($\widetilde{U}$) experienced by electrons differs from the applied potential ($U_{0}$) because of Coulomb screening. We can account for this difference using the Hartree equation

\begin{align}
    \widetilde{U}_{q} = U_{0q} - \frac{2 \pi e^{2}}{\varepsilon q} n_{q}
\end{align}

Where $n_{q}$ is the Fourier transform of the spatially varying particle density of a system described by Eqn.(1) in the main text with $U = \widetilde{U}$. Here $\varepsilon = 13$ is the dielectric constant of GaAs. The subscripts $q$ denote Fourier coefficients. The Hartree equation must be solved at a particular Fermi energy, $E_{F}$. In real space, the particle density is defined by

\begin{align}
    n(r) = \sum_{\substack{k,n \\ E_{k,n} < E_{F}}} | \psi_{k, n}(r) |^{2}
\end{align}

Where $\psi_{k,n}(r)$ are the Bloch functions obtained via numerical diagonalisation (see Methods). To solve the Hartree equation we assume that $\widetilde{U}$, like $U_{0}$, is sinusoidal, meaning that its only non-zero Fourier component is $\widetilde{U}_{g_{i}}$. Evaluating the Hartree equation at $q = g_{i}$ we obtain

\begin{align}\label{eqn:hartree}
    \widetilde{W} = W_{0} - \frac{2 \pi e^{2}}{\varepsilon g_{i}} \frac{n_{g_{i}}}{A_{cell}}
\end{align}

Where $A_{cell}$ is the unit cell area in real space and

\begin{align}
    n_{g_{i}} = \int_{cell} d^{2} r \ n(r) e^{i g_{i} \cdot r}
\end{align}

\begin{figure}
    \centering
    \includegraphics[width=0.55\textwidth]{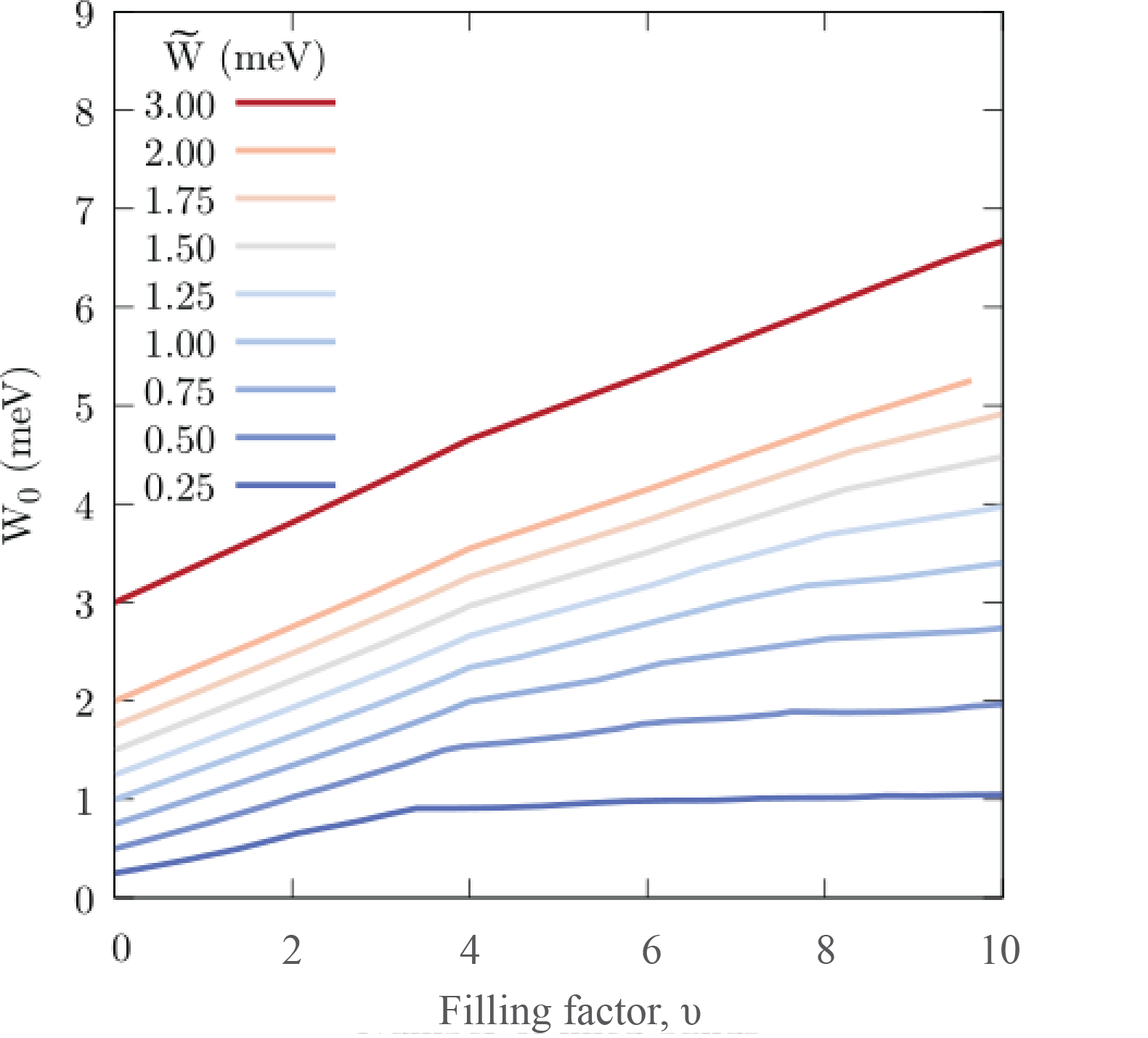}
    \caption{Effect of Coulomb screening on the potential amplitude. Each curve is the applied potential $W_{0}$ necessary to obtain a given self-consistent potential $\widetilde{W}$ at different electron densities. Electron density is given on the horizontal axis in units of the number of filled electron bands. Each curve corresponds to a different value of $\widetilde{W}$.}
    \label{fig:screenedW}
\end{figure}

Using Eqn.~\ref{eqn:hartree} we can find the strength of the applied potential $W_{0}$ necessary to obtain a self-consistent potential strength $\widetilde{W}$. The results of this calculation are presented in Fig.~\ref{fig:screenedW}. We remind that the parameter $W$ is related to the peak-to-peak width $U_{p-p}$ of the potential by $U_{p-p} = 9 W$.

\section{Estimation of modulation strength}

Before we can estimate the potential amplitude a few points must first be made clear.

\begin{enumerate}
    \item Potential amplitude decreases as electron density increases. For example, an imposed potential of $W_{0} = 3$ meV is reduced to $\widetilde{W} = 1$ meV when four energy bands are fully filled (at $n = 4 n_{0} = 9.2 \times 10^{10} \ \text{cm}^{-2}$). This is the result of Fig. \ref{fig:screenedW}.

    \item At large potential amplitudes each energy band is distinct, there is no overlap between bands. As the potential amplitude is reduced bands begin to overlap starting with the higher-energy bands first (see Figs. 1c-e in the main text).

    \item If the bands of interest (the graphene-like and kagome-like bands) are distinct for a given imposed potential, $W_{0}$, then there will be a `critical' density $n_{c}$ such that for all $n > n_{c}$ the energy bands are overlapping. This occurs because the potential strength decreases with density.

    \item When the energy bands are distinct there is only ever one kind of charge carrier present for a given electron density (either electron-like or hole-like). When the energy bands overlap both kinds of charge carrier can exist simultaneously. In the latter case no clear signature of the bands is expected in the Hall slope and experimentally this is signaled by a suppression of the Hall response.
    
\end{enumerate}

We use these points to estimate the size of the potential amplitude. %at $V_{TG} = -0.5$ V (Fig. ~\ref{fig:evolution}(b)) and $V_{TG} = - 3$ V (Fig.~\ref{fig:evolution}(g)). 
In particular, we identify the value of $V_{PG}$ which corresponds to $n_{c}$. At voltages larger than this $V_{PG}$ no Hall slope transitions are observed. We assign a value of $\widetilde{W}$ at this $V_{PG}$ using our band structure calculation: the band structure must give the correct number of Hall slope transitions below $n_{c}$.

Starting with Fig.~\ref{fig:evolution}(b) we note that three transitions in the Hall slope are observed below $V_{PG} = 0.87$ V. Above $V_{PG} = 0.87$ V there are no further transitions. We identify the point $V_{PG} = 0.87$ V with the density $n_{c}$ defined above. Using the notation defined in this section the band structure of Fig. 2(e) was computed for $\widetilde{W}(n_{c}) = 0.45$ meV. This band structure is consistent with the observed Hall slope transitions and with our estimation that the electron density is less than $n_{0} / 2 = 1 / A_{cell} = 1.15 \times 10^{10} \ \text{cm}^{-2}$ at the lowest value of $V_{PG}$ in Fig.~\ref{fig:evolution}(b) ($V_{PG} = 0.829$ V). Using Fig. \ref{fig:screenedW} we find that the imposed potential amplitude must be $W_{0} = 1$ meV.

We can repeat this analysis for the data in Fig. 3 and corresponding full set of maps in Fig.\ref{fig:evolution}. If we consider the strongest modulation Fig.~\ref{fig:evolution}(g) the lowest value of $V_{PG}$ corresponds to an electron density just above the first Dirac point. The first transition above this point corresponds to the second van Hove singularity ($V_{PG} = 0.927$ V). The next transition after this ($V_{PG} = 0.935$ V) is the top of the graphene-like bands. From the second transition point, the subsequent $0.024$ V is identified with the third energy band (the flat band). Above the third band there are two further transitions in Hall slope. Again, the difference, $\Delta V_{PG}$, between the top of the flat band and the second subsequent transition is $0.024$ V. Each of these transitions, and the spacings between them are consistent with a potential amplitude $\widetilde{W} = 1$ meV at complete filling of four energy bands. At lower fillings the value of $\widetilde{W}$ is larger (band structure shown in Fig. 3e). Using Fig. \ref{fig:screenedW} we find that $\widetilde{W}$ is $1.25$ meV when three bands are fully filled and $1.75$ meV when 2 bands are fully filled. Finally, the value of the imposed potential amplitude is $W_{0} = 3$ meV. Note that in Fig. 3(c) we plot the bandstructure for $\widetilde{W} = 1.5$ meV ($U_{p-p}=13.5$ meV) which corresponds to a filling in the flat band region.

\section{Estimation of disorder and mobility}

There are 3 kinds of disorder in our device.
(i) A short-range disorder related to the presence of impurities
in the host semiconductor. (ii) A short-range disorder related
to imperfections in nano-lithography process. (iii) A long-range disorder 
(puddles)  related to imperfections in nano-lithography process. 
The disorder of kind (i) can be neglected because the host semiconductor
is very clean, having a 
mobility of $\sim 500,000~\mathrm{cm^{2}/Vs}$ (at 2DEG density $n=2\times10^{11}~\mathrm{cm}^{-2}$). Disorder of kinds (ii) and (iii)
have been simulated numerically in detail in Ref. \onlinecite{Tkachenko15}. It
was shown that suppression of short wave-length harmonics due to the Poisson equation make
type (ii) less 
important;  puddle formation (type iii) is the most dangerous one, with the size of the 
puddles ranging from a few to several lattice periods. We thus concentrate on type (iii).
First, we assume that puddles result in a Lorentzian broadening of the
chemical potential

\begin{eqnarray}
\frac{\Gamma/2}{\pi(\delta \mu^2+\Gamma^2/4)} \ ,    
\end{eqnarray}

\begin{figure}[h!]
\centering
\includegraphics[width=0.6\textwidth]{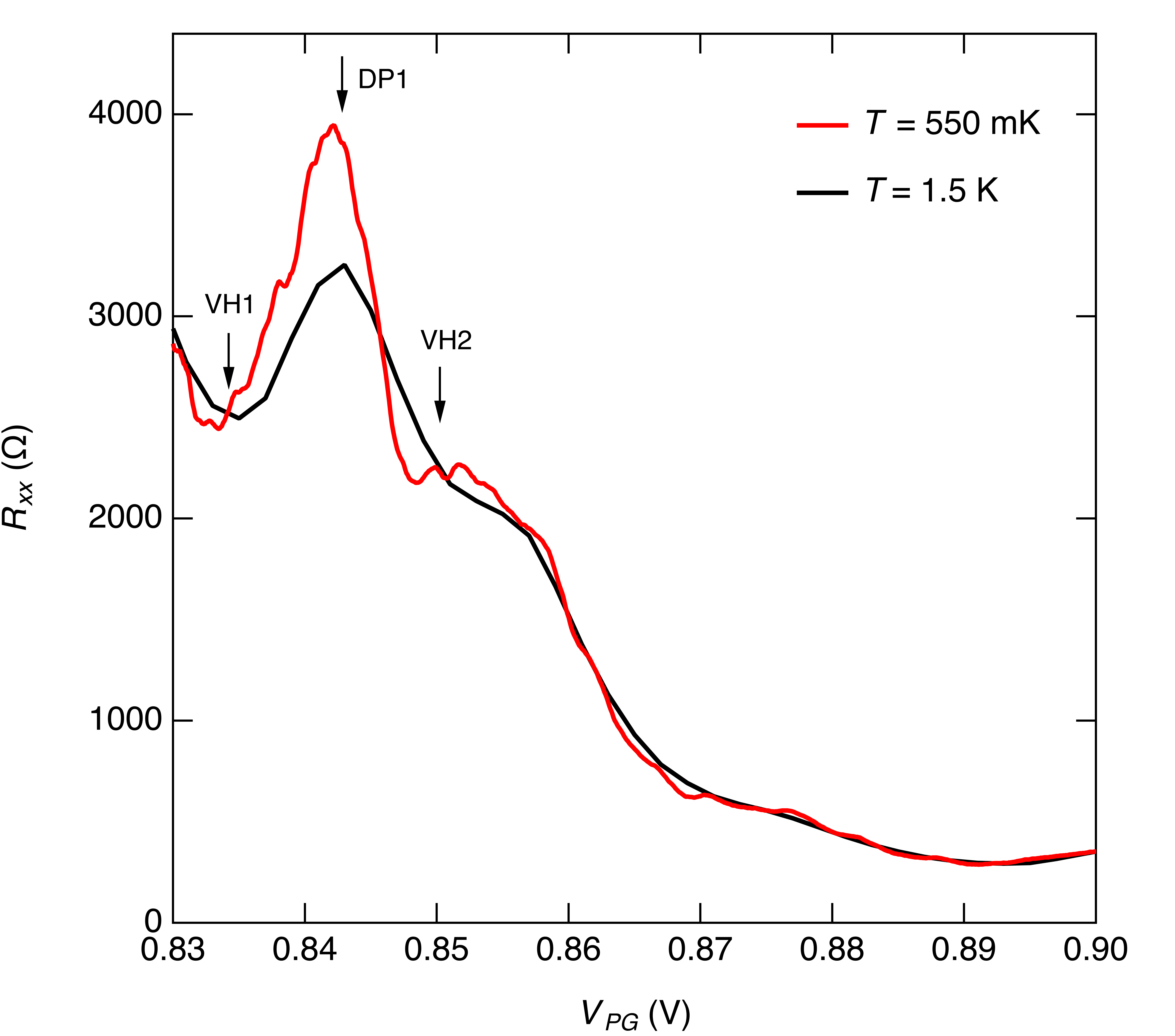}
\caption{$R_{xx}$ as a function of $V_{PG}$ at $V_{TG}=-0.5$ V. Two traces at different temperatures 1.5 K and 550 mK are plotted for comparison. The peak in $R_{xx}$ corresponds to the same Dirac peak in Fig. 2 in the main text. Black arrows indicate the positions the Dirac point and van Hove singularities from the Hall measurement.}
\label{fig:peakwidth}
\end{figure}

Using the width of the resistance peak at the first Dirac point DP1 we estimate the full-width-half-maximum $\Gamma$. Fig.~\ref{fig:peakwidth} shows this peak at 
two temperatures temperatures, $T = 550$ mK and $T=1.5$ K.
The $T=1.5$ K peak is slightly broader due to additional temperature broadening.
To estimate the disorder broadening $\Gamma$ we use the $T = 550$ mK data.
If we take the base of the peak to be the resistance values at VH1 and VH2 then the full-width-half-maximum at this temperature is $\delta V_{PG} = 7.9$ mV. The difference in $V_{PG}$ between VH1 and VH2 is $16$ mV. Thus, the full-width-half-maximum $\delta V_{PG}$ accounts for half of the change in electron density between VH1 and VH2.
To translate this to an energy scale we take half of the energy difference between VH1 and VH2 in the computed bandstructure of Fig. 2d. Doing this we find 
 
\begin{eqnarray}
  \Gamma = 0.2~meV 
\end{eqnarray}

Next, mobility is estimated using the standard equation 

\begin{align}
    \mu = \frac{1}{n_{eff} e \rho_{xx}}
\end{align}

Where $n_{eff}$ is the effective density of charge carriers and $\rho_{xx}$ is resistivity. Note that both the chemical potential and the mobility are denoted by the
same letter $\mu$, we hope that the meaning of $\mu$ will be clear from its context. 
To estimate mobility near van Hove singularities VH1 and VH2 we have to remember 
that the effective density
at the Dirac point is zero and the effective density
at the van Hove singularities is about
$n_e\approx n_0/4=5.7\times10^{9}~\mathrm{cm^{-2}}$.
The measured longitudinal resistance presented 
in Fig. 2(g) and in  Fig.~\ref{fig:peakwidth} is
about $2400~\Omega$; to obtain resistivity we must multiply by the
Van der Pauw contant $\frac{\pi}{\ln{2}}=4.53$, appropriate for a device 
with Van der Pauw geometry.
Hence, $\rho_{xx}=4.53 R_{xx} \approx 11,000 ~\Omega$.
The mobility close to VH1 and VH2 is about
 
\begin{eqnarray}
    \mu =100,000~\mathrm{cm^{2}/Vs}
\end{eqnarray}

This has to be compared with the mobility of a free electron gas
in GaAs at density $n_e=n_{eff}=5.7\times10^{9}~\mathrm{cm^{-2}}$
which is about $10,000~\mathrm{cm^{2}/Vs}$.

\section{Estimation of quantum capacitance}
\label{sec:cap}
We consider a simplified device setup with PG and 2DEG as the plates of a parallel plate capacitor. The total energy of the system consists of the energy of electric field ${\cal E}_E$ between the plates and energy of electrons ${\cal E}_e$ due to the Fermi motion.
The energy of electric field comes from the volume between PG and 2DEG
\begin{eqnarray}
  \label{Efield}
&&{\cal E}_E = \frac{\epsilon E^2}{8\pi}Ad=2\pi \sigma^2Ad/\epsilon
\end{eqnarray}
Here A is the area of the plate and $\sigma=en$ is the density of charge
in 2DEG. The value of electric field follows from Gauss law,
$E=4\pi\sigma/\epsilon$. Here we use CGS units.
We consider non-interacting electron, hence kinetic energy of all electrons is
\begin{eqnarray}
  \label{Eel1}
 {\cal E}_e=A\int_0^{\mu}\varepsilon \rho(\varepsilon)d\varepsilon \ ,
\end{eqnarray}
where $\rho$ is density of states.
Here we set the chemical potential zero at the bottom of the lowest band.
Hence, the total energy is
\begin{eqnarray}
  \label{Et1}
        {\cal E}_t={\cal E}_E+{\cal E}_e=2\pi \sigma^2Ad/\epsilon+
A\int_0^{\mu}\varepsilon \rho(\varepsilon)d\varepsilon \ .
\end{eqnarray}
The variation of energy at the variation of charge, $\delta\sigma=e\delta n$, is
\begin{eqnarray}
  \label{Et2}
  \delta {\cal E}_t=4\pi (Ad/\epsilon)   \sigma \delta\sigma +
A\mu\delta n
\end{eqnarray}
The chemical potential is determined by the condition
\begin{eqnarray}
  \label{mu}
n=\int_0^{\mu}\rho(\varepsilon)d\varepsilon \ .
\end{eqnarray}
When differentiating Eq.(\ref{Eel1}) we have in mind that
$\frac{\partial}{\partial n}=\frac{\partial\mu}{\partial n}\frac{\partial}{\partial \mu}=\frac{1}{\rho(\mu)}{\partial \mu}$.
  This is why the density of states disappears from Eq.(\ref{Et2}).
  To justify the cancellation we need to say that even within a gap there is
  some small density of states.
  On the one hand the variation of energy is (\ref{Et2}) and on the other hand it
  is equal to
  $V\delta Q=VA\delta\sigma$, where is the applied voltage. Hence the relation between voltage and
  electron number density is
  \begin{eqnarray}
    \label{Vd}
    V=\frac{4\pi d}{\epsilon} e n +\frac{\mu}{e}
  \end{eqnarray}
To find $V(n)$ one needs to solve this equation together with (\ref{mu}).   

%%%%%%%%%%%%%%%%%
\begin{figure}[h!]
 \centering
\includegraphics[width=0.8\textwidth]{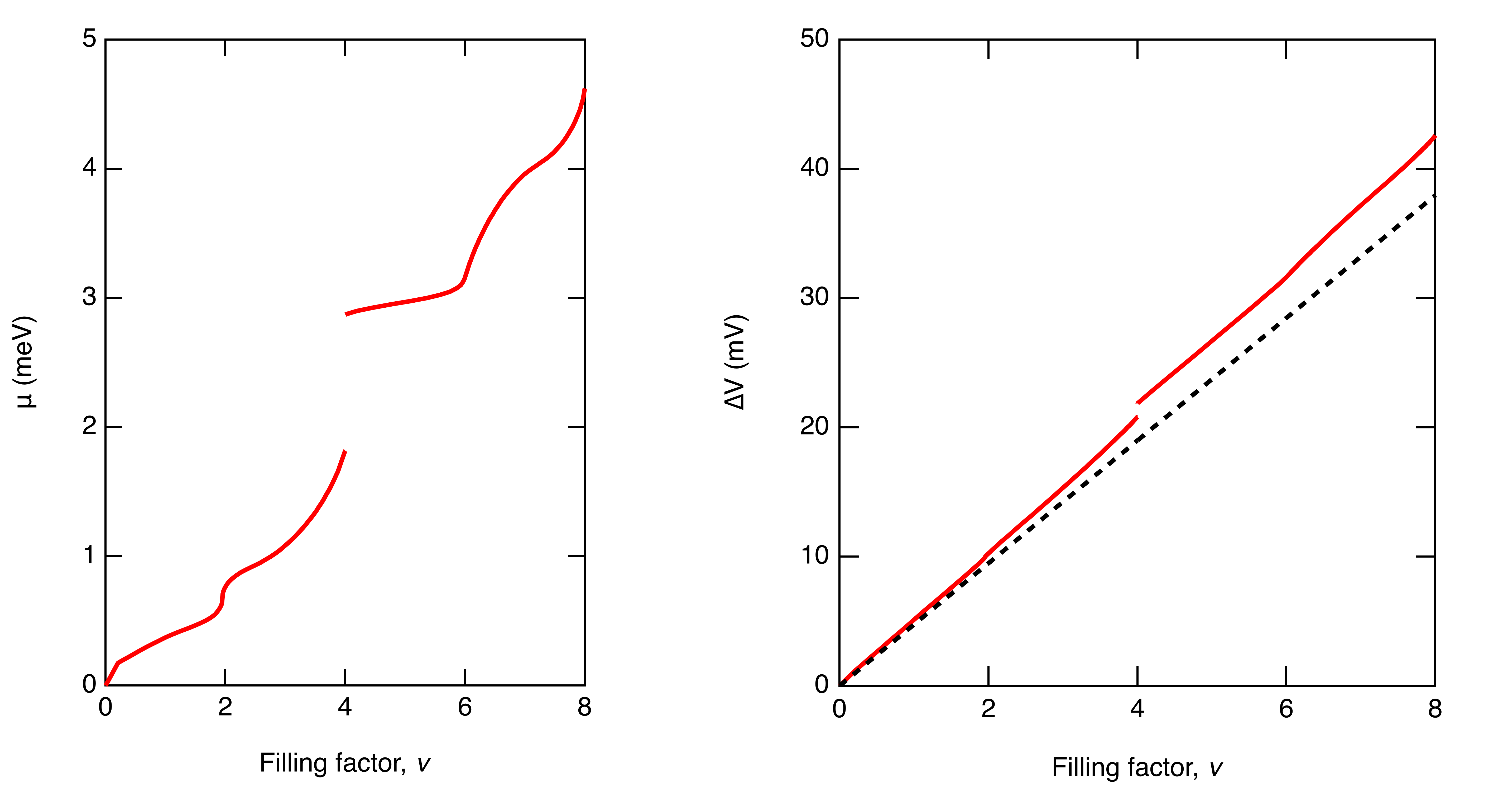}
\caption{(a) Chemical potential versus the band filling factor. A single band
  filling  $\Delta\nu =2$ corresponds to the electron density
  variation  $\Delta n = 2.31\times 10^{-10}cm^{-2}$.
  (b) Voltage versus the band filling factor. The red solid line represents
  the calculation with account of the quantum capacitance. The black dashed line
  is given by the simple classical formula.  
}
\label{fig:quantumcap}
\end{figure}
%%%%%%%%%%%%%%%%%

Consider the case of a modulation strength $U_{p-p}=13.5$meV (Fig.3(c) of main text) and $\mu(n)$ for this dispersion (Fig.\ref{fig:quantumcap}(a)), we can calculate $V(\nu)$ using Eq.~(\ref{Vd}). In Fig.~\ref{fig:quantumcap}(b) we plot the calculated $V(\nu)$ (equivalently $\Delta V$ in experiment) for $d = 25$ nm as the red solid trace.  At $\nu=4$ there is a tiny discontinuity $\sim 1$ mV due to the band gap. In comparison we also plot the simple
classical case  $V=\frac{4\pi d}{\epsilon} e n$ (black dashed line). 
The calculated average slop $\Delta V=10$meV per $\Delta \nu =2$ is smaller than the experimental value $\Delta V=24$meV because we use a simplified model of the capacitor. In the real device, the thickness of the 2DEG is $\sim 15$ nm, which gives the capacitor a much larger effective distance $d$.  The real electric field is also much more complicated because of the superlattice patterning. Nonetheless, this calculation shows that deviations from the straight line due to the quantum capacitance contribution are very small and overall the filling factor is still linear to gate voltage $V_{PG}$. If considering the real band size of $\Delta V = 24$ mV, the disorder-related discontinuity of $\sim 1$ mV caused by the band gap is $\sim 4\%$ of the band capacity.

\section{Loop Current Wigner Insulator}

\begin{figure}[h!]
  \includegraphics[width=0.2\textwidth]{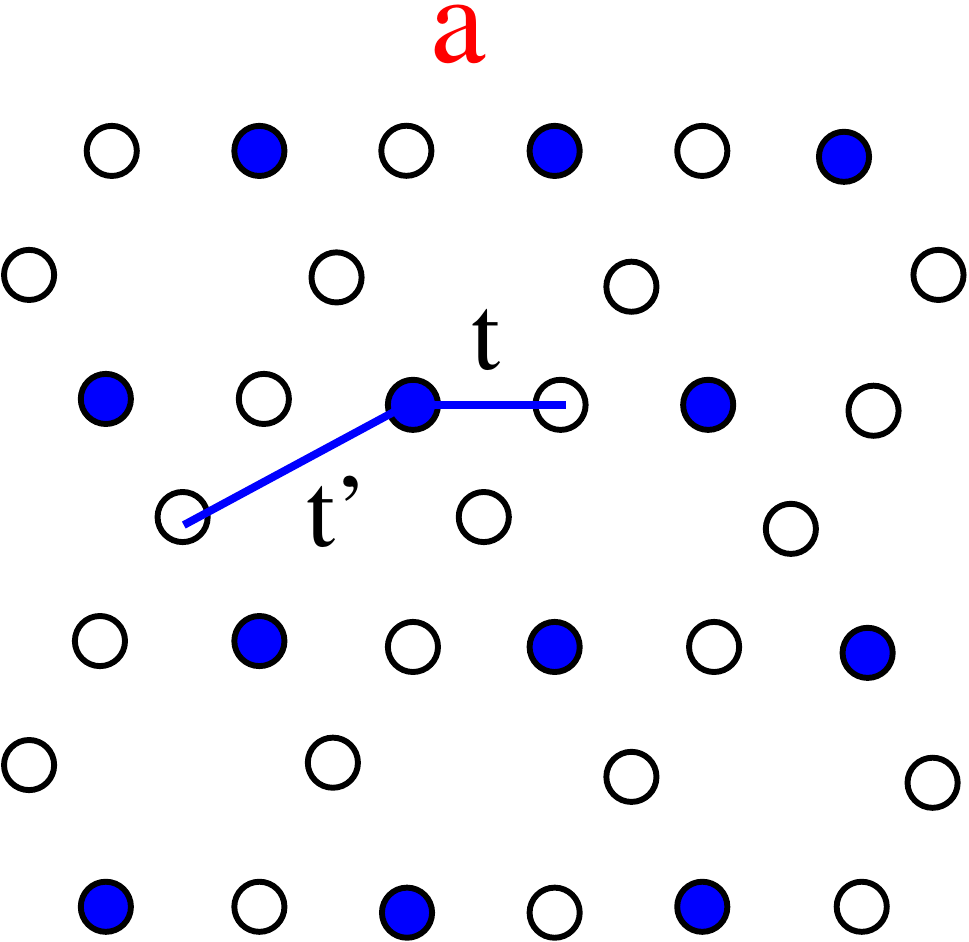}
  \hspace{30pt}
    \includegraphics[width=0.2\textwidth]{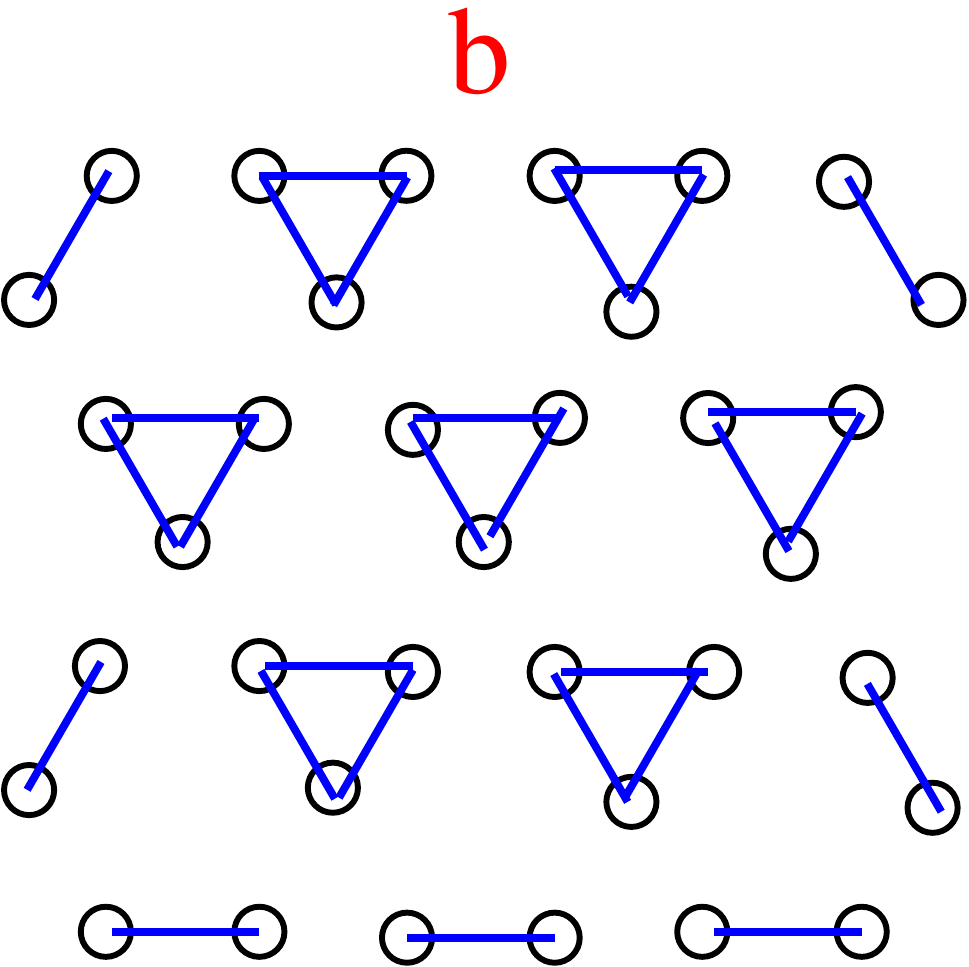}
  \hspace{30pt}
  \includegraphics[width=0.2\textwidth]{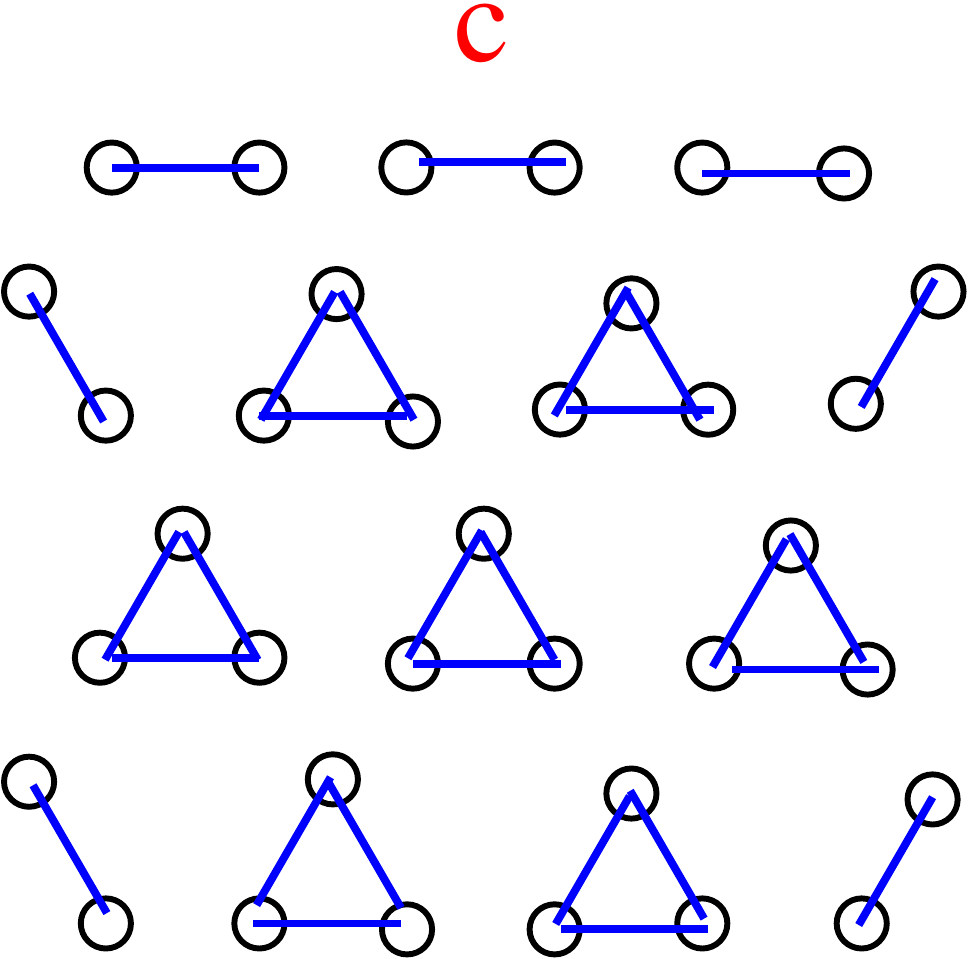}
  \caption{ a. 1/3 filling of the kagome lattice in a commensurate Winger insulator configuration with empty circles representing empty lattice sites and blue circles representing electrons. $t$ and $t^{\prime}$ indicate the nearest and next nearest neighbour hopping matrix elements respectively. b and c are two possible configurations of the loop current Wigner insulator with solid lines indicating the circulating current.
    }
\label{Wign}
\end{figure}

We observe the strong insulating behaviour at half filling of the flat kagome
band. This filling of the band corresponds to 1/3 filling of the kagome tight
binding lattice as it is shown in Fig.~\ref{Wign}a, where a blue dot corresponds to one
electron.
%%%%%%%%%%%%%%%%%
%%%%%%%%%%%%%%%%%
This configuration can be stabilised only by the long-range Coulomb repulsion.
Let us denote by $t$ the nearest site hopping matrix element and by $t^\prime$
the next nearest site hopping matrix element on the kagome lattice (Fig.~\ref{Wign}a).
The energy of the state Fig.~\ref{Wign}a can be lowered by allowing the
electron to hop around the triangle as shown in Fig.~\ref{Wign}b.
The delocalization within the triangle practically does not influence the
Coulomb energy, but decreases the kinetic energy.
The electron delocalization on a triangle leads to a current around
the triangle. This loop current is similar to the loop current 
suggested for cuprates~\cite{Varma97,Varma2020}.
Hence we arrive at the picture of Loop Current Wigner Insulator
that consists of a triangular lattice of loop currents.
There are two possible configurations of triangular loop current in a kagome lattice as
shown in panels b and c of Fig.~\ref{Wign}. Hence, lattice wise, the loop
current state is double degenerate.

The electron dispersion in non-interacting approximation has been
calculated and we can fit the dispersion using the tight binding model.
For the kagome bands the distance between the flat band and the kagome
Dirac point (DP2) is $3t$. Hence, comparing with the band structure we
conclude that for $U_{p-p}=13.5$ meV the nearest site hopping matrix element is $t\approx 0.59$ meV and for  $U_{p-p}=27$ meV it is $t\approx 0.53$ meV. Similarly one can fit the next nearest hopping: for $U_{p-p}=13.5$ meV the value is $t^\prime \approx -0.07$ meV and $U_{p-p}=27$ meV it is $t^\prime \sim +0.02$ meV.
Naturally $t^\prime \ll t$ and $t$ is not  very sensitive to $U_{p-p}$
within the reasonable range of the modulation. Hence we neglect $t^\prime$ in all estimates
except of the antiferromagnetic interaction of circulating currents which
arises only due to $t^\prime$.

The Coulomb interactions, the on-site Hubbard $U$, the nearest neighbours
interaction $V_1$ (distance L/2), and the next nearest neighbours interaction
$V_2$ (distance L) are
\begin{eqnarray}
  \label{UVV}
  &&U \approx \frac{e^2}{\varepsilon L/4}\left(1-\frac{1}
        {\sqrt{1+\left(\frac{8d}{L}\right)^2}}\right)\approx 3.5meV   \nonumber\\ 
        &&V_1 \approx \frac{e^2}{\varepsilon L/2}\left(1-\frac{1}
        {\sqrt{1+\left(\frac{4d}{L}\right)^2}}\right)\approx 1.0meV 
        \nonumber\\ 
  &&V_2\approx \frac{e^2}{\varepsilon L}\left(1-\frac{1}
        {\sqrt{1+\left(\frac{2d}{L}\right)^2}}\right)\approx 0.2meV 
  \end{eqnarray}
Here $L=100$ nm is the lattice spacing, $\epsilon=13$ is the dielectric
constant and $d\approx 35$nm is the effective distance to the gate that provides
screening via the image charge. $V_1$ and $V_2$  are the most important
parameters and their estimates are pretty simple and reliable. The estimate
for $U$ is rather crude, but it is confirmed by the numerical integration with
electron density.

%Fig.~\ref{tri} shows two neighbouring triangles, a and b.
%%%%%%%%%%%%%%%%%
\begin{figure}[h!]
  \includegraphics[width=0.25\textwidth]{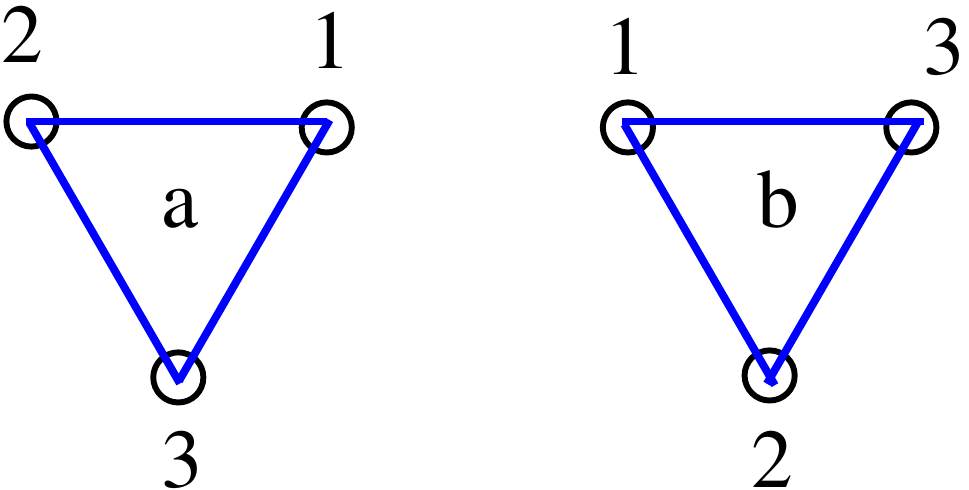}
  \caption{ Two neighbouring triangles a and b of the loop current Wigner insulator with sites on each triangle numbered.
    }
\label{tri}
\end{figure}
%%%%%%%%%%%%%%%%%
Consider first the single electron dynamics within a triangle.
There are three eigenstates
\begin{eqnarray}
  \label{wfs}
  &&\epsilon =2t\cos\frac{2\pi}{3}=-t, \ \ \ \ \ \
  a_+^\dag=\frac{1}{\sqrt{3}}\left[a_1^\dag +e^{2\pi i/3}a_2^\dag +e^{4\pi i/3}a_3^\dag\right]\nonumber\\
  &&\epsilon =2t\cos\left(-\frac{2\pi}{3}\right)=-t, \ \ \ \ 
  a_-^\dag=\frac{1}{\sqrt{3}}\left[a_1^\dag +e^{-2\pi i/3}a_2^\dag +e^{-4\pi i/3}a_3^\dag\right]\nonumber\\
  &&\epsilon =2t\cos(0)=2t, \ \ \ \ \ \ \ \ \ 
  a_0^\dag=\frac{1}{\sqrt{3}}\left[a_1^\dag +a_2^\dag +a_3^\dag\right]
\end{eqnarray}    
Here $a_i^\dag$ is the electron creation operator at the site i
of the triangle a.
The states for the b-triangle in Fig.~\ref{tri}, $b_+^\dag,b_-^\dag,b_0^\dag$,
are similar.

Since $ t > 0$ the ground state energy is $-t$, this is the gain in energy
due to electron delocalization. 
The ground state is double degenerate,
$a_\pm^\dag$. These states carry electric currents (loop currents).
Performing the Peierls substitution, $t \to t e^{ie\int {\bm A}\cdot {\bm dl}}$,
and expanding at small B we find the magnetic moment per triangle
\begin{eqnarray}
  \label{mu}
  \mu_\pm=\pm\frac{etL^2}{16}\approx \pm 10\mu_B  
\end{eqnarray}
The value corresponds to $L=100$ nm and $t=0.59$ meV.

To address the stability of the loop current state
one needs to consider virtual hopping of an electron from one triangle to
another. Here we follow the same approach as Anderson's superexchange
for a Mott insulator.
Hopping from one triangle to another creates two
electrons on a triangle. The two-electron states are
\begin{eqnarray}
  \label{2e}
  &&parallel \ spins\nonumber\\
  &&  |1\rangle =b_{+\uparrow}^\dag b_{-\uparrow}^\dag|0\rangle , \ \ \ \ E_1=-2t+V_1\nonumber\\
  &&  |2\rangle=b_{+\uparrow}^\dag b_{0\uparrow}^\dag|0\rangle , \ \ \ \ E_2=t+V_1\nonumber\\
  &&antiparallel \ spins\nonumber\\
  &&  |3\rangle=b_{+\uparrow}^\dag b_{+\downarrow}^\dag|0\rangle , \ \ \ \ E_3=-2t+2V_1/3+U/3\nonumber\\
  &&  |4\rangle=b_{+\uparrow}^\dag b_{-\downarrow}^\dag|0\rangle , \ \ \ \ E_4=-2t+2V_1/3+U/3\nonumber\\
  &&  |5\rangle=b_{+\uparrow}^\dag b_{0\downarrow}^\dag|0\rangle , \ \ \ \ E_5= t+2V_1/3+U/3
\end{eqnarray}
Let us consider the loop current state $b_{+\beta}^\dag a_{+\alpha}^\dag$,
where $\alpha$ and $\beta$ are spin indexes.
The energy of the state is
\begin{eqnarray}
  \label{E0}
  E_0=-2t+V_2
\end{eqnarray}
The hopping Hamiltonian $tb_{1\sigma}^\dag a_{1\sigma}$ leads to mixing of the zero
approximation wave function $b_+^\dag a_+^\dag $ with $b_+^\dag b_\alpha^\dag$,
where $\alpha=+,-,0$. There is also an equal contribution from $b \to a$
hopping. All in all this leads to the second order perturbation theory energy
shift $\delta E$
(Anderson mechanism).
\begin{eqnarray}
  \label{dE2}
&&parallel \ spins, \ \alpha=\beta: \ \ \ \ \ \ \ \ 
\delta E= -2(t/3)^2\left(\frac{1}{V_1-V_2}+\frac{1}{V_1-V_2+3t}  \right)\nonumber\\
&&antiparallel \ spins,\ \alpha \ne \beta: \ \
\delta E= -2(t/3)^2\left(\frac{2}{2V_1/3+U/3-V_2}+\frac{1}{2V_1/3+U/3+3t-V_2}  \right)
\end{eqnarray}
There are three conclusions from these equations\\
(i) The most ``dangerous'' term is that with the denominator $V_1-V_2$,
hence the stability of the state depends on the long-range interactions,
the on-site Hubbard U is of secondary importance.
For stability of the state one  needs $(t/3)/(V_1-V_2) \ll 1$, according to
our estimates  the value of the parameter is  $(t/3)/(V_1-V_2) \approx 0.25 $.\\
(ii) The spin parallel orientation has lower energy than the antiparallel one.
The energy difference is about 0.03 meV. Hence the spin dynamics are described
by ferromagnetic spin  Hamiltonian, $s_i=1/2$,
\begin{eqnarray}
  \label{Hs}
  && H_s=-J_ss_as_b\nonumber\\
  &&J_s \sim 0.06 ~\mathrm{meV}\sim 0.5 ~\mathrm{K}
  \end{eqnarray}\\
(iii) The energies (\ref{dE2}) are independent of the relative orientation
of loop currents, hence while at $T=0$ spins are ordered ferromagnetically,
the Ising orbital magnetic moments $\pm 10\mu_B$ remain fully disordered.

The energy splitting between different directions of the loop currents
arises only with account of the next nearest hopping $t^\prime$.
The second order correction to the energy dependent on the relative current
orientation arises due to the following sequences of hoppings,
$3a \to 1b$ and then back $2b \to 1a$,  also 
$1a \to 2b$ and then back $1b \to 3a$ (Fig.~\ref{tri}).
For simplicity consider only ferromagnetic orientation of spins and
account only for the virtual hoppings that have the minimal energy denominator, 
$\Delta E=V_1-V_2$. This gives the following energy shifts:
for the same direction of loop currents
$\delta E=\frac{2(t^{\prime}/3)^2}{V_1-V_2}$,
for opposite directions of loop currents
$\delta E=-\frac{4(t^{\prime}/3)^2}{V_1-V_2}$.
Hence the antiferromagnetic ordering of the loop currents is preferable
and the effective Hamiltonian of Ising moments $l_i=\pm 1$ is
\begin{eqnarray}
  \label{Hl}
  && H_l=J_ll_al_b\nonumber\\
  &&J_l \approx \frac{1}{3} \frac{(t^{\prime})^2}{V_1-V_2} \sim 0.002~\mathrm{meV} \sim 20~\mathrm{mK}
\end{eqnarray}
Since $J_l$  is very small the loop currents are practically always disordered
at $B=0$.
However, an application of an external out of plane magnetic field orders the
currents since each loop carries magnetic moment $\sim 10\mu_B$.
Electrons thermally excited over the gap to the ``conduction band'' provide
conductivity. They scatter from thermal fluctuations of the loop currents,
ordering of the currents by magnetic field reduces the fluctuations in this
results in the giant negative magnetoresistance.